\documentclass[11pt,showkeys,nofootinbib,notitlepage]{revtex4-1}

\usepackage{amsmath,amssymb,amsfonts,amsthm}
\usepackage{amsbsy} 
\usepackage{epsfig}
\usepackage{latexsym}
\input amssym.def
\input amssym.tex

\usepackage{hyperref}

\newcommand{\oarX}[1]{\href{http://arxiv.org/abs/#1}{{\ttfamily #1}}}
\newcommand{\arX}[1]{\href{http://arxiv.org/abs/#1}{{\ttfamily arXiv:#1}}}
\newcommand{\doin}[2]{\href{http://dx.doi.org/#1}{#2}}

\def\barr{\begin{array}}
\def\earr{\end{array}}
\def\half{\frac{1}{2}}
\def\ben{\begin{equation}}
\def\een{\end{equation}}
\def\bs{\begin{subequations}}
\def\es{\end{subequations}}
\def\bena{\begin{eqnarray}}
\def\eena{\end{eqnarray}}

\def\im{{\rm i}}

\def\M{\mathcal{M}}
\def\be{\begin{equation}}
\def\ee{\end{equation}}
\def\bes{\begin{eqnarray}}
\def\ees{\end{eqnarray}}

\begin{document}

\title{Inhomogeneous universe from group field theory condensate}

\author{Steffen Gielen}
\affiliation{School of Mathematical Sciences, University of Nottingham, University Park, Nottingham NG7 2RD, UK}
\email{steffen.gielen@nottingham.ac.uk}

\date{\today}


\begin{abstract}
One of the fundamental challenges for quantum cosmology is to explain the emergence of our macroscopic Universe from physics at the Planck scale. In the group field theory (GFT) approach to quantum gravity, such a macroscopic universe results from the formation of a ``condensate'' of fundamentally discrete degrees of freedom. It has been shown that the effective dynamics of such GFT condensates follows the classical Friedmann dynamics at late times, while avoiding the classical singularity by a bounce akin to the one of loop quantum cosmology (LQC). It was also shown how quantum fluctuations in a GFT condensate provide an initial power spectrum of volume fluctuations around exact homogeneity. Here we connect the results for quantum fluctuations in GFT to the usual formalism for cosmological perturbations within quantum field theory in curved spacetime. We consider a bouncing universe filled with a massless scalar field, in which perturbations are generated by vacuum fluctuations in the contracting phase. Matching conditions at the bounce are provided by working within LQC. We then compare the results to the GFT condensate scenario for quantum gravity with massless scalar matter. Here, instead, an initial quantum phase described by a GFT condensate generates initial scalar perturbations through quantum fluctuations. We show general agreement in the predictions of both approaches, suggesting that GFT condensates can provide a physical mechanism for the emergence of a slightly inhomogeneous universe from full quantum gravity.
\end{abstract}


\maketitle

\tableofcontents

\newpage
\section{Introduction}
On the largest observable scales, the structure of our Universe appears to show great simplicity: far from requiring the complications of full general relativity, it can be described by linear perturbations on a homogeneous and isotropic background spacetime. The statistical pattern of the perturbations can itself be fully described by a simple functional relation, leaving only a few free parameters whose observed values require deeper explanation in the physics of the early universe. This explanation is commonly assumed to be provided by inflation \cite{Baumann:2009ds}, with bouncing cosmologies of various types providing the most studied alternative \cite{bounceReview}. Inflation and many alternative approaches are framed within quantum field theory on curved spacetime, which assumes a semiclassical (continuum) spacetime on which quantum fields can be defined. Moreover, the approximate homogeneity of the universe usually needs to be assumed as a property of the initial state, globally or at least locally. Both assumptions certainly demand further fundamental explanation within quantum cosmology and, presumably, a theory of quantum gravity.\footnote{For instance, the trans-Planckian problem of inflation suggests the need to go beyond semiclassical spacetime \cite{TransPl}.} One suggestion for a new physical principle setting initial conditions for inflation within quantum cosmology, the no-boundary proposal \cite{noboundary}, has been discussed controversially \cite{controversy}; in general, the issue of a fundamental justification of the initial assumptions for inflation remains open. In bouncing scenarios, the need to avoid a big crunch/big bang singularity also requires non-standard assumptions, such as including effects of a putative theory of quantum gravity or exotic matter fields violating the Null Energy Condition. Without resolving the cosmological singularity, the transition from the contracting into the expanding phase introduces additional ambiguities into the physical predictions of bouncing cosmologies, since matching conditions at the bounce are required \cite{brandenfinelli}. (The need for matching conditions might be avoided by exploiting analyticity properties of general relativity in the complex time plane \cite{qprop}.) While theoretical cosmology is able to account for almost all current observations within a self-contained framework, this framework itself is waiting for a completion by deeper foundations within a theory of quantum gravity.

From the perspective of quantum gravity, being able to give a description of the Universe we observe is a highly nontrivial task. In many approaches, showing the emergence of an effective macroscopic continuum spacetime structure from different (``non-spatiotemporal'') fundamental degrees of freedom is often the first challenging step \cite{philo}, even before questions of homogeneity and isotropy or the existence of a linearised regime as in cosmology can be addressed. In this article, we focus on the group field theory (GFT) approach \cite{GFTreviews}, whose fundamental degrees of freedom are discrete ``atoms of geometry'', arranged in combinatorial structures and enriched with group-theoretic data corresponding to a discrete gravitational connection and discretised matter fields. The combination of combinatorial and lattice gauge theory methods places GFT at the intersection of matrix and tensor models \cite{matrixtensormodels} and loop quantum gravity \cite{LQG}, of which GFTs can be viewed as a type of second quantisation \cite{GFT2ndq}. In such discrete approaches, a continuum regime is typically reached as a thermodynamic limit associated to a phase transition; in GFT specifically, the proposal is that a macroscopic continuum spacetime emerges from a GFT condensate, a highly coherent configuration of a large number of ``atoms''. The geometric interpretation of such a condensate, in the limit where the number of atoms is very large, is that of a macroscopic, homogeneous continuum geometry \cite{GFTcosmo}. The dynamics of such condensates, described to leading order by the GFT analogue of the Gross-Pitaevskii equation in condensed matter physics, can then be translated into effective cosmological dynamics, i.e., effective Friedmann equations for the effective continuum geometry. Such Friedmann equations have been derived most successfully for models of quantum gravity coupled to a massless scalar field \cite{GFTfriedmann}, which is the model used, mostly for technical simplicity, in much of the quantum cosmology literature \cite{BlythIsham}, and in particular in loop quantum cosmology \cite{LQC}. General overviews on the connection of GFT to (quantum) cosmology can be found in \cite{GFCreview}. The results of \cite{GFTfriedmann} show that most of the features of the effective loop quantum cosmology (LQC) dynamics, in the ``improved dynamics'' prescription \cite{improdyn}, could be derived from the dynamics of GFT condensates, and hence from a full quantum gravity setting.\footnote{See \cite{LQGLQC} for alternative approaches towards connecting the effective LQC dynamics to full loop quantum gravity.}

While promising in showing how GFT could reproduce the correct semiclassical physics at late times while resolving cosmological singularities, these results were rather restricted in that they only applied to flat homogeneous and isotropic spacetimes. As a step towards a realistic cosmology, the inclusion of inhomogeneities is crucial; one can then also start addressing questions such as why the initial state of the Universe should have been homogeneous and isotropic to such good approximation. In inflation, the inhomogeneities observed in the cosmic microwave background arise from quantum fluctuations present in the inflaton due to the uncertainty principle. Similarly, as GFT is a quantum theory for gravity and matter, it too should provide a non-vanishing spectrum of primordial inhomogeneities purely from quantum fluctuations. This idea was made precise in \cite{GFCperp}; after defining GFT operators corresponding to fluctuations in the local volume element (i.e., scalar perturbations), it was shown how quantum fluctuations present in a GFT condensate give these local volume perturbations a non-vanishing power spectrum. It was moreover observed that the power spectrum for a certain variable characterising these perturbations was scale-invariant. Furthermore, it was outlined how these results could be extended to other observables, in particular perturbations in the local energy density of matter, again finding a non-vanishing, scale-invariant power spectrum in certain observables.

In this article, we connect these results in the framework of GFT condensate cosmology to the conventional formalism of scalar perturbations in cosmology. The work of \cite{GFCperp} left two important questions open: it did not explain how the power spectrum for GFT volume perturbations was to be converted into the usual gauge-invariant variables for observable scalar perturbations in cosmology; perhaps more importantly, the observed initial power spectrum of quantum fluctuations is a kinematical property of a condensate at a given time, but this initial spectrum of quantum fluctuations must then be converted into classical perturbations in the propagation, ``freeze out'' and amplification during the expansion of the universe, in order to lead to a potentially observable spectrum of classical perturbations. Since the dynamics of the expanding universe were not used, this second step was also missing in \cite{GFCperp}. The purpose of the present article is to fill these gaps, and show how the proposal of \cite{GFCperp} that cosmological perturbations are seeded by quantum fluctuations in a quantum gravity condensate is realised in practice. For mostly technical reasons, the only GFT models successfully used to define condensates describe quantum gravity coupled to a set of free, massless scalar fields, and we will also work in this setting. The massless scalar fields serve a dual role, both as matter sources for gravity generating scalar perturbations and also as relational harmonic coordinates \cite{harmonic}, circumventing the usual problem of defining coordinates from matter degrees of freedom in a fully background-independent context.

The plan of this article is as follows. In section \ref{sec2}, we discuss standard cosmological perturbation theory in a flat homogeneous and isotropic universe, where the matter content is a massless scalar field. We envisage a modified ``matter bounce'' scenario \cite{matterbounce} in which cosmological perturbations arise from quantum fluctuations inside the Hubble radius in the contracting phase. As the universe contracts, these fluctuations leave the Hubble radius and are converted into classical perturbations. The issue of how to match the spectrum of these perturbations across the bounce into the expanding phase is addressed by working within the framework of LQC, which provides a non-singular bounce by incorporating quantum geometry corrections (coming from loop quantum gravity) into the classical Friedmann dynamics. We closely follow \cite{edmatterbounce}, where a dust-dominated universe was discussed in LQC, in matching classical low-curvature regions to high-curvature regions around the bounce. Propagation through the bounce then provides a primordial spectrum of perturbations, which eventually re-enter the Hubble radius in the expanding phase. In our case where the matter content is formed by massless scalars, the resulting power spectrum does not match actual CMB observations, but we exemplify this case to illustrate what one would expect from the standard formalism within quantum field theory on curved spacetime. In section \ref{sec3}, we then illustrate the implications of the alternative GFT condensate scenario: here the semiclassical low-curvature universe described within standard perturbation theory is preceded by a deep quantum-gravity phase, in which perturbations originate as quantum fluctuations in the GFT field (i.e., in a field theory for quantum gravity and matter). Again, this sets an initial power spectrum for scalar perturbations that is then propagated into the semiclassical universe.\footnote{Since the effective background geometry for the condensate phase undergoes a bounce, the GFT scenario resolves the classical singularity as well, but unlike other bounce scenarios such as the one of section \ref{sec2} it does not suggest that cosmological perturbations are generated in the preceding collapse phase.} Conceptually, this GFT condensate scenario bears some resemblance to string gas cosmology \cite{stringGas} where perturbations generated in a quantum-gravity phase (there given by a string gas) then also propagate according to the standard Einstein equations. We show general agreement between the physical predictions of the semiclassical framework and of the GFT condensate scenario, and thus consistency of the quantum gravity results of GFT with the usual semiclassical approach. Our results suggest that GFT condensates can provide a physical mechanism for the emergence of not only an exactly homogeneous, but also a slightly inhomogeneous and hence more realistic universe from full quantum gravity. We conclude with a discussion of the results in section \ref{sec4}.

\section{Bouncing Universe with a Massless Scalar}
\label{sec2}

Consider a spatially flat homogeneous and isotropic universe with metric
\ben
{\rm d}s^2 = -{\rm d}t^2 + a(t)^2\,h_{ij}(x){\rm d}x^i{\rm d}x^j
\een
given in terms of proper time $t$ and a flat 3-metric $h_{ij}$. In the conventions we use in this article, coordinates $t$ and $x^i$ carry dimensions of length (in the convention $c=1$) while $h_{ij}$ and $a(t)$ are dimensionless. We will frequently switch between the variable $t$ and conformal time $\eta$ defined by
\ben
\eta = \int\limits^t \frac{{\rm d} t'}{a(t')}\,,
\een
which also has dimensions of length. In order to describe a non-singular bounce obtained by quantum gravity effects, we will assume that the scale factor $a(t)$ follows not the usual Friedmann equations of classical cosmology but instead the effective equations of LQC. These effective equations can be derived from the full quantum cosmology formalism of LQC if the quantum state for the cosmological background is suitably semiclassical \cite{taveras}, i.e., they capture the leading quantum gravity corrections. They take the form
\ben
\left(\frac{1}{a}\frac{{\rm d}a}{{\rm d}t}\right)^2 = \frac{8\pi G}{3}\rho\left(1-\frac{\rho}{\rho_c}\right)\,,\quad \frac{{\rm d}\rho}{{\rm d}t}+3\frac{1}{a}\frac{{\rm d}a}{{\rm d}t}(\rho+P)=0\,.
\label{LQCfriedmann}
\een
Here $\rho$ is the energy density and $P$ the pressure of matter, $G$ is Newton's constant and $\rho_c$ is a critical density usually taken to be Planckian, which characterises the scale at which quantum geometry effects kick in. Clearly once the matter density reaches $\rho=\rho_c$, a bounce occurs. Notice that the energy-momentum conservation equation is identical to the usual one in general relativity.
\\With matter taken to be a massless scalar, we have $P=\rho$ and hence
\ben
\rho(t) = \frac{M^4}{a(t)^6}\,,
\label{energydens}
\een
with $M$ a constant with dimensions of mass (now also assuming $\hbar=1$); interestingly, one can then also find the exact solution for $a(t)$, given by \cite{edmatterbounce}
\ben
a(t)=\frac{M^{2/3}}{\rho_c^{1/6}}\left(1+24\pi G\, t^2\rho_c\right)^{1/6}
\een
where we fix the freedom to rescale $a(t)$ by an arbitrary constant to obtain an expression that is dimensionless and well-defined in the limit $\rho_c\rightarrow\infty$ in which LQC corrections disappear. This solution is clearly non-singular, but turns into a singular solution $a(t)\sim t^{1/3}$ as $\rho_c\rightarrow\infty$. Such a simple solution in terms of elementary functions cannot be found when $a(t)$ is expressed in terms of conformal time $\eta$, so that we will use proper time whenever the full LQC solution is needed.

\begin{figure}[htp]
\includegraphics[scale=0.85]{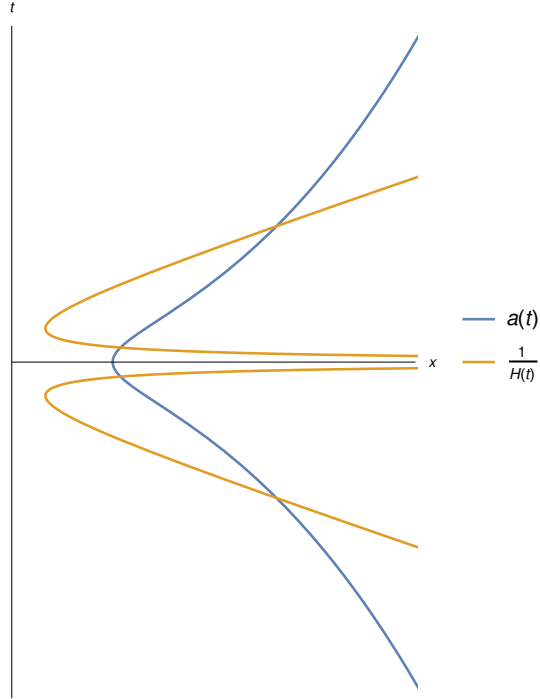}
\caption{Scale factor $a(t)$, governing the stretching of a given initial perturbation mode, vs. evolution of the Hubble radius which sets the local scale on which perturbations ``feel'' the effect of spacetime curvature.}
\label{fig1}
\end{figure}
To visualise the propagation of cosmological perturbations in this LQC bounce geometry, it is now instructive to plot both the scale factor $a(t)$ and the size of the Hubble radius $\frac{1}{H(t)}\equiv\frac{a(t)}{{\rm d}a/{\rm d}t}$ as functions of cosmic time $t$ \cite{BrandenbergerIntro}. We show this in Fig.~\ref{fig1} for a universe in LQC filled with a massless scalar. The figure illustrates how a given perturbation mode generated in the far past inside the Hubble radius shrinks with the contraction of the universe but eventually exits the Hubble radius which contracts faster. Likewise, in the expanding phase, the Hubble radius expands faster than the spacetime geometry as a whole and a mode can re-enter at late times.

Let us now describe the dynamics of linearised scalar perturbations in this cosmological scenario. Such perturbations are most conveniently described in terms of the gauge-invariant and canonically normalised Mukhanov-Sasaki variable $v$ \cite{sasakimukhanov}. Including LQC corrections \cite{LQCpert}, $v$ satisfies the differential equation
\ben
\frac{{ \rm d}^2 v}{{\rm d}\eta^2} + \left(k^2\left(1-\frac{2\rho}{\rho_c}\right)-\frac{1}{z}\frac{{\rm d}^2 z}{{\rm d}\eta^2}\right)v =0
\label{mukhsas}
\een
where we work in Fourier space so that $v(\eta,k)$ depends both on conformal time $\eta$ and a wavenumber $k$, and the variable $z$ is given by
\ben
z(\eta)=\sqrt{\frac{3}{4\pi G}}\,\frac{a(\eta)}{\sqrt{1-\frac{\rho(\eta)}{\rho_c}}}\,.
\een
Solving (\ref{mukhsas}) exactly is rather difficult. However, in the cosmological scenario we are interested in, we can perform several justified approximations. Since we assume cosmological perturbations are generated in the far past where $\rho\ll\rho_c$, LQC correction terms may initially be dropped. The equation for $v$ is then simply (using $z(\eta)\propto \sqrt{-\eta}$ in conformal time)
\ben
\frac{{ \rm d}^2 v}{{\rm d}\eta^2} + \left(k^2+\frac{1}{4\eta^2}\right)v =0
\label{muksakeq}
\een
which has the general solution
\ben
v(\eta,k)=A(k)\sqrt{-k\eta}\,J_0(-k\eta)+B(k)\sqrt{-k\eta}\,Y_0(-k\eta)
\een
where $J_0$ and $Y_0$ are Bessel functions of the first and second kind. One is now interested in two kinds of asymptotic regimes: first the long wavelength/late time limit $|k\eta|\ll 1$ for which
\ben
v(\eta,k) \sim \sqrt{-k\eta}\left[A(k)+\frac{2B(k)}{\pi}\left(\gamma+\log\left(-\frac{k\eta}{2}\right)\right)\right]\,,
\label{longwave}
\een
where $\gamma\approx 0.577$ is the Euler-Mascheroni constant, and then the short wavelength/early time limit $|k\eta|\gg 1$ for which
\ben
v(\eta,k) \sim e^{-\im k\eta}\frac{1-\im}{2\sqrt{\pi}}\left(A(k)-\im B(k)\right)+e^{\im k\eta}\frac{1+\im}{2\sqrt{\pi}}\left(A(k)+\im B(k)\right)\,.
\een
In this discussion, ``early times'' refers to the far past long before the bounce while ``late times'' is still in the contracting phase, but close to the bounce before the Planckian regime is reached. These explicit expressions for the asymptotic behaviour of the mode functions $v(\eta,k)$ confirm the picture obtained by ignoring either the $k^2$ or the $\frac{1}{\eta^2}$ term in (\ref{muksakeq}): at early enough times, any mode with a given $k$ must have been inside the Hubble radius, where the mode simply oscillates as a plane wave. At late times, the mode is outside the Hubble radius and grows or shrinks as a power law in $\eta$. 

The quantity one is interested in when describing the amplitude of observable scalar perturbations is not $v$ but the ``curvature perturbation in comoving gauge'' $\zeta=v/z$. In terms of $\zeta$, the amplitude of oscillating modes grows in the contracting phase, whereas the amplitude of modes outside the Hubble radius is almost constant, changing only logarithmically in conformal time.

The challenge for fundamental cosmology is to provide a physical mechanism that fixes the functional form of $A(k)$ or $B(k)$ in the initial state. In this section, we look at the treatment in standard cosmological scenarios, where one assumes that the quantum field described by $v(\eta,k)$ started out in the vacuum state in which the normalisation of the modes is
\ben
|v(\eta,k)|\stackrel{|k\eta|\rightarrow -\infty}{\sim} \frac{1}{\sqrt{2k}}e^{-\im k\eta}\quad \Rightarrow \; v(\eta,k)=\frac{\sqrt{-\pi\eta}}{2}H_0^{(1)}(-k\eta)\,.
\label{hankel}
\een
$H_0^{(1)}$ is the Hankel function of the first kind, defined by $H_0^{(1)}=J_0+\im Y_0$.

As the universe contracts, each mode will eventually leave the Hubble radius and ``freeze out'',
\ben
v(\eta,k) \stackrel{|k\eta|\ll 1}{\sim} \sqrt{\frac{-\eta}{\pi}}\left(\im\gamma+\frac{\pi}{2}+\im\log\left(-\frac{k\eta}{2}\right)\right)\,.
\label{lowk}
\een
Eventually, the contraction of the universe reaches the regime where LQC corrections including the critical density $\rho_c$ become important. In order to propagate the mode $v(\eta,k)$ across the non-singular bounce, we now need to take these corrections into account and match (\ref{lowk}) to a solution in the Planckian regime. Notice that, as expected, without LQC corrections the curvature perturbation $\zeta$ would diverge as $|k\eta|\rightarrow 0$, and linear perturbation theory would break down. This is why a non-singular cosmological scenario is needed in which $\zeta$ can remain finite throughout the evolution.

Reintroducing LQC corrections into the Mukhanov-Sasaki equation
\ben
\frac{{ \rm d}^2 v}{{\rm d}\eta^2} + \left(\left(1-\frac{2\rho}{\rho_c}\right)k^2-\frac{1}{z}\frac{{\rm d}^2 z}{{\rm d}\eta^2}\right)v =0\,,
\een
we still do not expect to be able to solve this equation in general. Since we are however now only interested in the limit $|k\eta|\ll 1$ for a given mode, we can discard the $k^2$ term and write down the general solution (here and in the following we follow closely the discussion of \cite{edmatterbounce} for a dust-dominated matter bounce in LQC)
\ben
v_{{\rm LQC}}(\eta,k)=B_1(k)z(\eta)+B_2(k)z(\eta)\int\limits^\eta \frac{{\rm d}\eta'}{z(\eta')^2}+O(k^2\eta^2)\,,
\label{vlqc}
\een
where $B_1(k)$ and $B_2(k)$ are arbitrary functions of $k$. This expression is a solution both in standard cosmology and with LQC corrections, but the functional form of $z$ is different in both cases. Namely, for our universe filled with a massless scalar in LQC, we have
\ben
a(t)=\frac{M^{2/3}}{\rho_c^{1/6}}\left(1+24\pi G\, t^2\rho_c\right)^{1/6}\,,\quad z(t)=\sqrt{\frac{3}{4\pi G}}\frac{a(t)}{\sqrt{1-\frac{\rho(t)}{\rho_c}}}\equiv-\frac{a(t)^4}{4\sqrt{2}\pi G\,M^2\,t}
\een
where the last equality follows from inserting the explicit form of $a(t)$. These expressions are given in proper time $t$, not conformal time $\eta$. However, we can calculate
\ben
\int\limits^\eta \frac{{\rm d}\eta'}{z(\eta')^2} \equiv \int\limits^t \frac{{\rm d}t'}{a(t')z(t')^2} = -\frac{4\pi G\,t\sqrt{\rho_c}}{3M^2}\left(1+24\pi G\, t^2\rho_c\right)^{-1/2}+\frac{2\sqrt{\pi G}}{3\sqrt{3}M^2}{\rm ArSinh}\left(\sqrt{24 \pi G\,\rho_c}\,t\right)+C
\label{integral}
\een
where $C$ is an integration constant to be chosen by the requirement that the LQC mode solution $v_{{\rm LQC}}$ can be matched to the mode solution (\ref{lowk}) computed in the contracting phase in the regime where LQC corrections are no longer relevant, $\sqrt{24\pi G\rho_c}t\ll -1$. In this limit, (\ref{integral}) becomes
\ben
\int\limits^t \frac{{\rm d}t'}{a(t')z(t')^2}  \rightarrow C-\frac{\sqrt{2\pi G}}{3\sqrt{3}\,M^2}\left(\log\left(-\sqrt{96 \pi G \,\rho_c}\,t\right)-1\right)
\een
or, written in conformal time $\eta$,
\ben
\int\limits^\eta \frac{{\rm d}\eta'}{z(\eta')^2}  \rightarrow C-\frac{\sqrt{2\pi G}}{3\sqrt{3}\,M^2}\left(\log\left(16M\left(\frac{2\pi G}{3}\right)^{3/4}\sqrt{\rho_c}(-\eta)^{3/2}\right)-1\right)
\een
which suggests that a useful choice is
\ben
C:=\frac{\sqrt{2\pi G}}{3\sqrt{3}\,M^2}\left(\log\left(\left(\frac{2\pi G}{3}\right)^{3/4}32\sqrt{2\rho_c}\,M\,k^{-3/2} \right)-1\right)\,.
\een
The sub-Planckian limit $\sqrt{24\pi G\rho_c}t\ll -1$ can then be taken, yielding
\ben
\int\limits^\eta \frac{{\rm d}\eta'}{z(\eta')^2}  \rightarrow -\frac{\sqrt{\pi G}}{\sqrt{6}M^2}\log\left(-\frac{k\eta}{2}\right)\,.
\een
In summary, we computed an approximate solution of the LQC corrected Mukhanov-Sasaki equations in the limit of low wavenumber $|k\eta|\ll 1$, which away from the Planck-scale regime and before the bounce approaches
\ben
v(\eta,k)\sim B_1(k)\left(\frac{6}{\pi G}\right)^{1/4}M\sqrt{-\eta}-B_2(k)\left(\frac{\pi G}{6}\right)^{1/4}\frac{\sqrt{-\eta}}{M}\log\left(-\frac{k\eta}{2}\right)\,.
\een
Matching this solution to the low wavenumber regime (\ref{lowk}) of the perturbations in the collapsing phase, we can now fix
\ben
B_1(k)=\left(\frac{G}{6\pi}\right)^{1/4}\frac{2\im\gamma+\pi}{2M}\,,\quad B_2(k)=-\im\left(\frac{6}{\pi G}\right)^{1/4}\frac{M}{\sqrt{\pi}}\,.
\label{coefficients}
\een
Notice that for the case of a universe filled with a massless scalar field, these coefficients are $k$-independent (but we included a $k$-dependence into the integration constant $C$). Because of the nontrivial background geometry, there is now mode mixing across the bounce. This can be seen by now considering the limit $\sqrt{24\pi G\rho_c}\,t\gg 1$ of the approximate LQC mode solution, which corresponds to the sub-Planckian regime after the bounce. In this limit, we have
\ben
\int\limits^\eta \frac{{\rm d}\eta'}{z(\eta')^2}  \rightarrow \frac{\sqrt{\pi G}}{\sqrt{6}M^2}\log\left(\frac{k\eta}{2}\right)+\frac{2\sqrt{2 \pi G}}{3\sqrt{3}M^2}\left[\log\left(\left(\frac{2\pi G}{3}\right)^{3/4} 32\sqrt{2\rho_c}\,Mk^{-3/2}\right)-1\right]
\label{largetlimit}
\een
where we reinstate the explicit expression for $C$ for clarity. We see that this second mode now contributes not only to the logarithmically growing mode, but also to the constant mode in the expanding universe. Concretely, putting (\ref{vlqc}), (\ref{coefficients}) and (\ref{largetlimit}) together, we have
\ben
v(\eta,k) \sim\sqrt{\frac{\eta}{\pi}}\left(\im\gamma+\frac{\pi}{2}-\im\log\left(\frac{k\eta}{2}\right)-\frac{4}{3}\im\left[\log\left(\left(\frac{2\pi G}{3}\right)^{3/4} 32\sqrt{2\rho_c}\,Mk^{-3/2}\right)-1\right]\right)
\een
in the expanding phase (and away from the Planckian regime in which LQC is relevant); notice the nontrivial last term which arose from the interaction of this mode with the bouncing background.

As we already mentioned, the quantity of observational interest is $\zeta=v/z$. Using $z(\eta)=M(6/\pi G)^{1/4}\sqrt{\eta}$ away from the LQC regime, we finally find
\ben
\zeta(\eta,k)= \frac{1}{M}\left(\frac{G}{6\pi}\right)^{1/4}\left(\im\gamma+\frac{\pi}{2}-\im\log\left(\frac{k\eta}{2}\right)-\frac{4}{3}\im\left[\log\left(\left(\frac{2\pi G}{3}\right)^{3/4} 32\sqrt{2\rho_c}\,Mk^{-3/2}\right)-1\right]\right)
\een
in the expanding phase. Notice that $\zeta$ is almost constant during the expansion of the universe, with only a logarithmic dependence on $k\eta$ which we shall ignore in the following (this contribution is small since $k\eta_*\sim 1$ at conformal time $\eta_*$ when the mode re-enters the Hubble radius). The resulting expression for the scalar power spectrum to be observed at late times is then
\ben
\Delta_\zeta^2(k)=\frac{k^3}{2\pi^2}|\zeta(\eta_*,k)|^2 \approx\frac{4\sqrt{2}k^3}{9\pi^2 M^2}\sqrt{\frac{G}{3\pi}}\log\left(\left(\frac{2\pi G}{3}\right)^{3/4} 32\sqrt{2\rho_c}\,Mk^{-3/2}\right)^2
\een
if we assume that the argument of the logarithm is of order $m_{{\rm Pl}}^{1/2}Mk^{-3/2}\gg 1$ for the modes of interest, i.e., the matter energy density given by $M$ is sufficiently non-negligible. In this limit, the dominant contribution to the power spectrum comes from the mode mixing at the LQC bounce. If this is not the case, the logarithm would be replaced by an $O(1)$ numerical coefficient. The prefactor of the amplitude is of order
\ben
\frac{4\sqrt{2}k^3}{9\pi^2 M^2}\sqrt{\frac{G}{3\pi}} \sim 0.02 \frac{k^3}{M^2 m_{{\rm Pl}}}\ll 0.02
\label{classiAmpl}
\een
where the last inequality is again the assumption that $M$ contributes a sufficiently large energy density for the modes of interest. In any case, $M$ is a free parameter in this model that could be fixed to match any observed amplitude of scalar perturbations. However, the resulting spectral index $n_s=4$ is fixed and clearly incompatible with our observations of the CMB.

\section{Universe with Massless Scalar from GFT Condensate}
\label{sec3}

We now consider a scenario in which a universe with the same matter content ``emerges'' in a framework for quantum gravity. This discussion requires providing some background regarding the approach we are working in, which is what we will do in this section.

The group field theory (GFT) approach to quantum gravity posits that a large semiclassical universe emerges from a ``condensate'' of fundamental quantum gravity degrees of freedom or ``atoms of space'' \cite{GFTcosmo, GFCreview}. For the convenience of readers mainly interested in (classical and quantum) cosmology, we summarise the salient features of GFT inasmuch as the application to cosmology is concerned. For general reviews of GFT as an approach to quantum gravity, see \cite{GFTreviews}.

A general GFT is defined in terms of a (complex) scalar field $\varphi$ on an abstract group manifold (not to be thought of as spacetime), with dynamics governed by an action
\ben
S[\varphi,\bar\varphi]=-\int {\rm d}^4 g\;{\rm d}^4\phi\;\bar\varphi(g_I,\phi^J)\mathcal{K}\varphi(g_I,\phi^J)+\mathcal{V}[\varphi,\bar\varphi]\,.
\label{akshn}
\een
In the application to inhomogeneous cosmology \cite{GFCperp}, the domain space of the field $\varphi$ is chosen to be ${\rm SU}(2)^4\times\mathbb{R}^4$, where the ${\rm SU}(2)$ arguments are denoted by $g_I$ and the real-valued arguments by $\phi^J$. $\mathcal{K}$ is a kinetic operator which generally contains derivatives with respect to both $g$ and $\phi$ variables. $\mathcal{V}$ is the interaction term consisting of terms of order $\varphi^3$ and higher; we will not specify the concrete form of $\mathcal{V}$ in what follows.

A concrete choice of $\mathcal{K}$ and $\mathcal{V}$ is usually made by demanding that the amplitudes associated to Feynman graphs of the theory equal the amplitudes for a given {\em spin foam model}, i.e., a discrete quantum gravity path integral. That is, in the perturbative expansion
\ben
\mathcal{Z}\equiv\int \mathcal{D}\varphi\,\mathcal{D}\bar\varphi\;e^{-S[\varphi,\bar\varphi]}=\sum_{\Gamma} \frac{\prod_i\lambda_i^{V_i}}{{\rm sym}[\Gamma]}A[\Gamma]
\label{GFTexpansion}
\een
of the GFT path integral, each Feynman graph (or vacuum bubble) $\Gamma$ represents a discrete spacetime history with associated (purely combinatorial) amplitude $A[\Gamma]$, weighted with a prefactor determined by the GFT coupling constants $\lambda_i$. (\ref{GFTexpansion}) generalises the corresponding expansion of the partition function for matrix models, which generates a sum over discrete two-dimensional Riemannian surfaces, i.e., a sum-over-histories for quantum gravity in two dimensions \cite{matrixtensormodels}, to higher dimensions. It was shown in \cite{BCGFT} how amplitudes for the Barrett-Crane spin foam model for four-dimensional quantum gravity \cite{barrettcrane} could be obtained from such a construction for a suitable choice of GFT action; later it was realised that any prescription  for a spin foam amplitude $A[\Gamma]$ (within a class of models of interest for quantum gravity) could be obtained from a GFT \cite{rovellireis}. In this way, the partition function of a GFT corresponds to a sum over topologies and spacetime histories, for gravity and matter, where each history itself is still discrete and contains a finite number of degrees of freedom. The main technical challenges, as for the lower-dimensional case of matrix models, are controlling the unwieldy sum over  $\Gamma$ and performing a continuum limit. Both challenges arise because the perturbative regime in which one truncates to simple $\Gamma$ does not correspond to relevant continuum physics.

The correspondence with discrete quantum gravity path integrals usually leads to a suitable bare action $S[\varphi,\bar\varphi]$, which must be augmented by quantum corrections under renormalisation. Radiative corrections then usually generate additional terms in $\mathcal{V}$ beyond those initially considered, and in particular derivative terms in $\mathcal{K}$ even if one has started from a trivial (ultralocal) quadratic action in which $\mathcal{K}$ was a pure number \cite{GFTrenorm}.

The amplitudes $A[\Gamma]$ can be expressed as sums or integrals over a suitable convolution of propagators, in the usual way. These sums or integrals are over variables corresponding to the arguments of the GFT field, i.e., in our case, elements of ${\rm SU}(2)^4\times\mathbb{R}^4$ or suitable ``dual'' spaces. The ``momentum'' representation for ${\rm SU}(2)$ results in a sum over irreducible representations (spins) associated to edges of the associated Feynman graph. The $\mathbb{R}^4$ labels $\phi^J$ or their dual momentum variables are associated to vertices of the graph. The interpretation of these variables is as discrete gravitational and matter degrees of freedom associated to each discrete spacetime history one is summing over in (\ref{GFTexpansion}). The ${\rm SU}(2)$ valued variables $g_I$ are interpreted as holonomies or parallel transports of an ${\rm SU}(2)$ connection (Ashtekar-Barbero connection \cite{AshtekarBarbero}) for gravity, while the real labels $\phi_J$ represent the values of matter (scalar) fields associated to each vertex in the graph. In this sense, choosing ${\rm SU}(2)^4\times\mathbb{R}^4$ as the domain space for a GFT field naturally leads to a proposal for a quantum theory for gravity coupled to four massless scalar fields.

Most of the work on GFT condensate cosmology applies to a general class of GFT models without specification of the forms of $\mathcal{K}$ and $\mathcal{V}$, also in light of the existence of a significant number of models in the literature with no clear consensus about the preference for one or the other as candidates for four-dimensional quantum gravity. (The class of models is somewhat restricted in that we have made a specific choice for ${\rm SU}(2)$ as the gauge group of gravity, a choice that could be generalised as well \cite{GFTcosmo}.) Within this class of GFT models one then uses an effective field theory approach in which the kinetic term is expanded in derivatives \cite{GFTfriedmann,GFCperp,lioritizhang},
\ben
\mathcal{K}=\mathcal{K}^0+\tilde{\mathcal{K}}^1\frac{\partial^2}{\partial(\phi^0)^2}+\mathcal{K}^1\sum_{i=1}^3\frac{\partial^2}{\partial(\phi^i)^2}+\ldots
\label{Kexpansion}
\een
where $\ldots$ includes fourth and higher derivatives. This expansion assumes that the GFT action is invariant under (constant) shifts in $\phi^I$, parity/time-reversal $\phi^I\mapsto -\phi^I$,
and rotations $\phi^i\mapsto {O^i}_j\phi^j$ where ${O^i}_j\in {\rm O}(3)$ and $i,j=1,2,3$. These symmetries are all symmetries of free, massless scalar fields, which is the type of matter we are interested in here. As a consequence, the coefficients $\mathcal{K}^i$ in this expansion can still contain derivatives with respect to the $g_I$ variables, but no explicit dependence on $\phi^I$ is allowed.

With (\ref{Kexpansion}), the equation of motion resulting from variation of (\ref{akshn}) is
\ben
\left(\mathcal{K}^0+\tilde{\mathcal{K}}^1\frac{\partial^2}{\partial(\phi^0)^2}+\mathcal{K}^1\sum_{i=1}^3\frac{\partial^2}{\partial(\phi^i)^2}+\ldots\right)\varphi(g_I,\phi^J)-\frac{\delta\mathcal{V}[\varphi,\bar\varphi]}{\delta\bar\varphi(g_I,\phi^J)}=0\,.
\label{EOM}
\een 
One now assumes the formation of a GFT condensate in which the quantum field $\hat\varphi$ acquires a nonvanishing expectation value,
\ben
\langle\hat\varphi(g_I,\phi^J)\rangle=\sigma(g_I,\phi^J)\,,
\een
as a candidate non-trivial configuration corresponding to a macroscopic semiclassical geometry (the ``vacuum'' $\varphi=0$ corresponds to no spacetime at all, in the interpretation of the Feynman expansion as a sum over discrete spacetime histories). In the simplest mean-field approximation, the field configuration $\sigma(g_I,\phi^J)$ should satisfy the classical equation of motion (\ref{EOM}). This is analogous to imposing the Gross-Pitaevskii equation on a condensate wavefunction $\Psi(\vec{x},t)$ in the usual treatment of a Bose-Einstein condensate in condensed matter physics, and an approximation to using the full quantum effective action instead (which would presumably contain an infinite number of terms).

The mean-field approximation is typically valid when interactions are sufficiently weak. This means that, for self-consistency, there must be a regime in which (\ref{EOM}) is approximately solved if one only considers the first kinetic term,
\ben
\left(\mathcal{K}^0+\tilde{\mathcal{K}}^1\frac{\partial^2}{\partial(\phi^0)^2}+\mathcal{K}^1\sum_{i=1}^3\frac{\partial^2}{\partial(\phi^i)^2}+\ldots\right)\sigma(g_I,\phi^J)= 0\,.
\label{simpleGP}
\een
The equation is now linear and explicit solutions can be found easily. This last approximation, perhaps the most drastic one, is not strictly necessary and the potential $\mathcal{V}$ has been included into studies of the effective dynamics of GFT condensates for some models \cite{KCLgroup}. We will assume (\ref{simpleGP}) mostly for technical simplicity. Since the potential contains higher powers of $\sigma$ and $\bar\sigma$ and the total particle number scales as $|\sigma(g_I,\phi^J)|^2$, interactions become dominant once the condensate reaches a certain size. The approximation in which interactions neglected hence corresponds to a mesoscopic regime in which the condensate has formed but not grown beyond a critical size \cite{GFTfriedmann}. There is in general no particle number conservation in these models; indeed the need to describe a realistic expanding universe generally leads to a preference for models with ``instabilities'' and exponential growth in the particle number, as we will see next.\footnote{This instability implies that the weak-coupling approximation breaks down after a finite time. In the cosmological scenario discussed below, this breakdown is not problematic since the quantum gravity GFT phase only describes an initial stage of the Universe in which perturbations are generated, which does not need to last very long.}

Explicit solutions to the linearised equation can be straightforwardly obtained by expanding functions on four copies of ${\rm SU}(2)$ into irreducible representations (Peter-Weyl expansion). For simplicity, one can restrict the mean field to only contain isotropic modes, characterised by only a single spin $j$; geometrically this assumption corresponds to a sum over discrete spacetimes that are constructed only out of equilateral tetrahedra. This restriction, while again not strictly necessary, turns out to be sufficient to capture the dynamics of homogeneous, isotropic universes, as one might have expected. With it, the mean field $\sigma$ is of the general form
\ben
\sigma(g_I,\phi^J)=\sum_{j\in \frac{\mathbb{N}_0}{2}} \sigma_j(\phi^J)\,{\bf D}^j(g_I)\,.
\een
All dependence on $g_I$ is now in the fixed functions ${\bf D}^j(g_I)$, obtained from four Wigner $D$-matrices which are contracted with suitable intertwiners. The expansion can be generalised to include modes with different $j$ labels for the four arguments of $\sigma$ \cite{anisopaper}.

The mode expansion of $\sigma$ into representation labels allows a decoupling of (\ref{simpleGP}) into independent equations for each $j$, at least in the simple linearised approximation that we consider. These equations are of the form (we also drop subleading higher derivative contributions from now on)
\ben
\left(-B_j+A_j \frac{\partial^2}{\partial(\phi^0)^2}+C_j \sum_{i=1}^3\frac{\partial^2}{\partial(\phi^i)^2}\right)\sigma_j(\phi^J)=0\,.
\label{jGP}
\een
Concretely, in the standard case in which $\mathcal{K}^0$, $\tilde{\mathcal{K}}^1$ and $\mathcal{K}^1$ are functions of Laplace-Beltrami operators with respect to the ${\rm SU}(2)$ variables $g_I$, the coefficients $A_j$, $B_j$ and $C_j$ are defined by
\ben
\mathcal{K}^0{\bf D}^j(g_I) =: -B_j{\bf D}^j(g_I)\,,\quad \tilde{\mathcal{K}}^1{\bf D}^j(g_I) =: A_j{\bf D}^j(g_I)\,,\quad \mathcal{K}^1{\bf D}^j(g_I) =: C_j{\bf D}^j(g_I)
\een
and each action of a Laplace-Beltrami operator on one of the $g_I$ variables would contribute an eigenvalue $-j(j+1)$. Expanding $\sigma$ in Fourier modes with respect to the $\phi^i$ coordinates, one can then write down a complete set of solutions to (\ref{jGP}), namely
\ben
\sigma_j^{K_i}(\phi^J)=\exp\left(\im K_i\phi^i\right)\left\{\alpha_j^+\exp\left(\sqrt{\frac{B_j+C_j K^2}{A_j}}\phi^0\right) + \alpha_j^-\exp\left(-\sqrt{\frac{B_j+C_j K^2}{A_j}}\phi^0\right)\right\}\,,
\label{sigmasolution}
\een
where $\alpha_j^+$ and $\alpha_j^-$ are arbitrary constants. 

The motivation for coupling four massless scalars to gravity in GFT (as opposed to the single scalar field we considered in section \ref{sec2}) is the dual role played by these scalar fields: as matter fields driving the expansion of the universe and sourcing cosmological scalar perturbations, and as labels that can be used to define local coordinates. Since nothing in the definition of GFT makes reference to any background structure that could be used to define a coordinate system, any identification of coordinates must be {\em relational}, i.e., use physical degrees of freedom whose values serve as coordinates at least locally. Massless scalar fields are perhaps the simplest possibility for such a relational coordinate system; including them circumvents the --in general very difficult-- problem of constructing coordinates from other degrees of freedom \cite{DeWitt,GHM}.

\subsection{Homogeneous Universe}

Keeping in mind that the $\phi^J$ serve as coordinates in space and time, a spatially homogeneous mean field now corresponds to the Fourier mode $\vec{K}=0$ in the general solution (\ref{sigmasolution}), i.e.,
\ben
\sigma_j(\phi^J)\equiv\sigma_j^{0}(\phi^0)=\alpha_j^+\exp\left(\sqrt{\frac{B_j}{A_j}}\phi^0\right) + \alpha_j^-\exp\left(-\sqrt{\frac{B_j}{A_j}}\phi^0\right)
\label{homosol}
\een
which is only a function of ``scalar field time'' $\phi^0$. In the following, we will assume that the condensate mean field takes this homogeneous form. This fine-tuned assumption about the initial state, while resulting in a homogeneous (background) universe, needs to be justified by further work on the possible initial conditions for GFT condensates, and on the stability behaviour of solutions to (\ref{jGP}). Our philosophy is to work with this simple class of condensate states and extract predictions for the (non-vanishing) resulting power spectrum of cosmological perturbations. One can certainly add perturbations to (\ref{homosol}), which would be analogous to phonons in real Bose-Einstein condensates; these would correspond to one form of perturbations of exact homogeneity, namely perturbations already in the background ``classical'' spacetime.\footnote{To include perturbations, one could also take an approach in which the mean field is taken to be constant within a ``patch'' for the $\phi^i$ coordinates, but where different patches do not necessarily have the same mean field (and thus, not the same effective homogeneous geometry). This approach corresponds to implementing the separate universe approach \cite{sepUniv} in GFT condensate cosmology \cite{GFTsepUniv}. We will not consider GFT corrections to the evolution equations for perturbations, as in \cite{GFTsepUniv}, but assume these are propagated by the classical Mukhanov-Sasaki equations.} In this paper, we are interested in an exactly homogeneous background configuration assumed to describe the primordial universe, with vacuum fluctuations that are to be converted into classical inhomogeneities in a later stage.

One can now extract an effective cosmological dynamics for such GFT condensates by considering the GFT operators corresponding to geometric volume observables. These can be defined straightforwardly by observing that $|\sigma_j(\phi^J)|^2$ corresponds to a local particle number density for quanta of spin $j$; with the input from loop quantum gravity that the local volume is a given function of the spins $j$, a local volume element is then given by \cite{GFCperp}
\ben
V(\phi^J) = \sum_j V_j |\sigma_j(\phi^J)|^2 \equiv \sum_{\{j_I\}} V_{\{j_I\}} \langle \hat\varphi^\dagger_{\{j_I\}}(\phi^J)\hat\varphi_{\{j_I\}}(\phi^J) \rangle
\label{volumeel}
\een
where $V_j$ is the eigenvalue of the loop quantum gravity volume operator for an elementary tetrahedron whose faces are all associated to a spin value (area) $j$. For large $j$, one has $V_j \sim V_{{\rm Pl}}j^{3/2}$, with deviations for small $j$ (see, e.g., \cite{bohrsommerf} for a detailed analysis of the spectrum). The last equality in (\ref{volumeel}) relates the specific expression valid for an isotropic mean field $\sigma_j(\phi^J)$ to the more general expression valid for an arbitrary state, in which case the sum runs over all combinations $\{j_I\}=\{j_1,\ldots,j_4\}$ of four spins. We will only consider states defined by an isotropic and homogeneous mean field of the form (\ref{homosol}).

One can then compute the evolution of local volume elements as a function of scalar field time $\phi^0$. The exact result depends, of course, on the choice of initial parameters $\alpha_j^+$ and $\alpha_j^-$; however, general statements can be made. In particular, for GFT models in which the ratio $\frac{B_j}{A_j}$ has a (positive) maximum for a given $j=j_0$, apart from the fine-tuned cases $\alpha_{j_0}^+=0$ or $\alpha_{j_0}^-=0$ {\em any} solution asymptotically satisfies \cite{lowj}
\ben
V(\phi^J) \stackrel{\phi\rightarrow \pm\infty}{\rightarrow} V_{j_0} |\alpha_{j_0}^\pm|^2\exp\left(\pm 2\sqrt{\frac{B_{j_0}}{A_{j_0}}}\phi^0\right)\,.
\label{volumeeq}
\een
In this case, for almost any initial state the spin $j=j_0$ will eventually dominate over all others and this domination is achieved exponentially fast. The evolution of the total volume then depends on the growth of the number of particles of spin $j_0$, given by the exponential factor appearing in (\ref{volumeeq}). Interestingly, this behaviour of the volume reproduces the classical solution for a universe filled with a massless scalar field, given by
\ben
V(\phi^0) = V_0 \exp\left(\sqrt{12\pi G}\phi^0\right)\,,
\label{classvol}
\een
if one identifies $\frac{B_{j_0}}{A_{j_0}}=:3\pi G$ with the low energy Newton's constant $G$. Showing this compatibility of the effective dynamics of GFT condensates with classical Friedmann cosmology was the first main result of \cite{GFTfriedmann}. One can say much more, though: it is easy to show that, except for cases of fine-tuned initial conditions, the volume elements $V(\phi^J)$ never go through zero throughout the evolution of a condensate. Notice also that the asymptotic behaviour (\ref{volumeeq}) shows a universe growing exponentially both to the past and the future, whereas the classical solution has a singularity as $V\rightarrow 0$ in the far past (due to our choice of scalar field time $\phi^0$, this singularity corresponds to $\phi^0\rightarrow -\infty$ even though it is reached in finite proper time). This classical singularity is hence resolved by GFT condensates. The last important result of \cite{GFTfriedmann} was that the corrections to classical Friedmann dynamics that appear when analysing certain specifically chosen mean field solutions in detail can be matched to the corrections to effective Friedmann equations that one sees in LQC (as in (\ref{LQCfriedmann})). Similar derivations of effective LQC-like dynamics from GFT condensates have also been given in the context of toy models \cite{Toy} and, more recently, in a different analysis of GFT using Hamiltonian methods \cite{GFThamilt}.

\subsection{Volume Perturbations from Quantum Fluctuations}
\label{volumeperp}

We have seen how GFT condensates can reproduce the correct semiclassical physics at low curvature, provide a mechanism for singularity resolution and connect to what was previously done in the context of LQC from a full quantum gravity setting. All of these are non-trivial results providing some evidence for the claim that GFT condensates might explain the emergence of a semiclassical universe from quantum gravity; but the setting of flat homogeneous, isotropic cosmology is only the very first approximation to situations one might be interested in in cosmology. The inclusion of cosmological perturbations into the formalism, in particular, is crucial. In this section we review the mechanism for the generation of cosmological perturbations in GFT condensates, and connect it to the previous analysis in the semiclassical LQC setting. More details and discussion can be found in \cite{GFCperp}.

The basic idea is conceptually rather similar to what is done in standard cosmology; namely, one obtains a non-vanishing power spectrum for scalar perturbations from the quantum fluctuations present in the initial quantum state that the Universe is in. Here, this is not a state for perturbations (defined, e.g., through the Mukhanov-Sasaki variable $v$ above) on some semiclassical background spacetime, but it is a state in the full GFT for gravity and matter, which is the key conceptual and technical difference to the standard treatment.

Following (\ref{volumeel}), let us define a local volume fluctuation operator given by
\ben
\delta \hat{V}(\phi^J) = \hat{V}(\phi^J) - \langle\hat{V}(\phi^J)\rangle\,,\quad \hat{V}(\phi^J)\equiv \sum_{\{j_I\}} V_{\{j_I\}} \hat\varphi^\dagger_{\{j_I\}}(\phi^J)\hat\varphi_{\{j_I\}}(\phi^J)
\een
and consider its two-point function at equal times,
\ben
\langle \delta \hat{V}(\phi^0,\phi^i)\delta \hat{V}(\phi^0,\phi'^i)\rangle = \sum_{\{j_I,j'_I\}} V_{\{j_I\}}  V_{\{j'_I\}} [\hat\varphi_{\{j_I\}}(\phi^0,\phi^i),\hat\varphi^\dagger_{\{j'_I\}}(\phi^0,\phi'^i)]\,\langle\hat\varphi^\dagger_{\{j_I\}}(\phi^0,\phi^i)\rangle\langle\hat\varphi_{\{j_I\}}(\phi^0,\phi'^i)\rangle
\label{correlator}
\een
where we again assume the mean-field approximation, i.e., that all normal-ordered $n$-point functions can be evaluated by replacing them with products of the corresponding one-point functions; the only non-trivial contribution to (\ref{correlator}) can then arise from the passage from a general $n$-point function to a normal-ordered one, which introduces the commutator of $\hat\varphi$ and $\hat\varphi^\dagger$. We have pulled this commutator out of the expectation value assuming that it is a c-number. Evaluating the two-point function then requires an explicit form for this commutator. There are two possible choices that have been discussed in the literature, with different arguments for deriving them: the first one is a ``timeless'' commutator algebra given by (e.g., \cite{GFTfriedmann,GFCperp})
\ben
[\hat\varphi_{\{j_I\}}(\phi^0,\phi^i),\hat\varphi^\dagger_{\{j'_I\}}(\phi'^0,\phi'^i)] = \delta_{j_I,j'_I}\,\delta(\phi^0-\phi'^0)\delta(\phi^i-\phi'^i)\,.
\label{commut1}
\een
Here all arguments of the GFT fields are treated as parameters on a configuration space, and the operators do not evolve in time, even in the Heisenberg picture. (One could in principle add a time dependence with respect to some background time $t$ and then require $\frac{\partial\hat\varphi}{\partial t}=\frac{\partial\hat\varphi^\dagger}{\partial t}=0$.) This picture is closest to that of usual canonical quantum gravity, where the Hamiltonian needs to vanish due to diffeomorphism symmetry and hence time evolution is ``frozen''.
\\Alternatively, one may prefer a commutator algebra closer to that of standard quantum field theory, in which one sees the fields as evolving with respect to the scalar field clock time $\phi^0$. This is a ``deparametrised'' formalism in which one of the degrees of freedom of the theory has been promoted to a time parameter. One can then perform a Legendre transformation of the action, introduce canonical momenta, and derive {\em equal-time} commutation relations
\ben
[\hat\varphi_{\{j_I\}}(\phi^0,\phi^i),\hat\varphi^\dagger_{\{j'_I\}}(\phi^0,\phi'^i)] = \delta_{j_I,j'_I}\,\delta(\phi^i-\phi'^i)\,,
\label{commut2}
\een
analogous to those introduced for GFT in \cite{Toy,GFThamilt}. In the following, we will assume (\ref{commut2}) which is closer to our philosophy of using $\phi^0$ as a ``clock'' with respect to which the evolution of the universe is defined. If (\ref{commut1}) is used instead, the difference in what follows would be that because of the delta function divergence in $\phi^0$ one cannot consider correlation functions of $\delta \hat{V}(\phi^J)$ but of a regularised quantity obtained by smearing $\delta \hat{V}(\phi^J)$ over an infinitesimal duration. Indeed, defining
\ben
\delta \hat{V}_{{\rm reg}}(\phi^J):=\int\limits_{\phi^0-\epsilon}^{\phi^0+\epsilon} {\rm d}\Phi^0\;\delta\hat{V}(\Phi^0,\phi^i)
\een
and evaluating the two-point function of $\delta \hat{V}_{{\rm reg}}$ would remove the factor $\delta(0)$ and lead to physical conclusions analogous to the ones reached for the volume perturbation variable $\delta \hat{V}$ and (\ref{commut2}). The choice between (\ref{commut1}) and (\ref{commut2}) is not very important when volume perturbations are considered.

A more subtle issue is the choice of commutators involving derivatives of fields with respect to $\phi^0$, as will be required in section \ref{gaugeinvperp}. Here, the equal-time commutation relation (\ref{commut2}) would not be sufficient to determine commutators between derived fields such as $\frac{\partial\hat\varphi}{\partial \phi^0}$ and $\frac{\partial\hat\varphi^\dagger}{\partial \phi^0}$, for example, and additional assumptions are needed. On the other hand, assuming (\ref{commut1}) would imply that, e.g.,
\ben
\left[\frac{\partial\hat\varphi_{\{j_I\}}}{\partial\phi^0}(\phi^0,\phi^i),\hat\varphi^\dagger_{\{j'_I\}}(\phi'^0,\phi'^i)\right] = \delta_{j_I,j'_I}\,\delta'(\phi^0-\phi'^0)\delta(\phi^i-\phi'^i)\,.
\label{commut3}
\een
Smearing this commutator over an infinitesimal $\phi^0$ period then gives zero for the right-hand side, as $\int_{-\epsilon}^{\epsilon}{\rm d}x\;\delta'(x)=\delta(\epsilon)-\delta(-\epsilon)=0$. For general consistency with this case, we will require in the following that all commutators involving $\phi^0$ derivatives of fields vanish. 

With (\ref{commut2}), the correlator (\ref{correlator}) becomes
\ben
\langle \delta \hat{V}(\phi^0,\phi^i)\delta \hat{V}(\phi^0,\phi'^i)\rangle = \delta(\phi^i-\phi'^i)\sum_{j} V^2_{j}  |\sigma_{j}(\phi^0)|^2
\een
where we again assume a homogeneous and isotropic mean field configuration (independent of $\phi^i$ and with only a single spin $j$ in the sum). The assumption of isotropy is again made for simplicity, but the assumption of homogeneity is crucial; it implies that the right-hand side, apart from the overall delta function, only depends on the scalar clock reading $\phi^0$. This result can be trivially converted into Fourier space, yielding \cite{GFCperp}
\ben
\langle \delta \hat{V}(\phi^0,K^i)\delta \hat{V}(\phi^0,K'^i)\rangle = (2\pi)^3 \delta(K^i+K'^i)\sum_{j} V^2_{j}  |\sigma_{j}(\phi^0)|^2
\label{volumeperps}
\een
where $K^i$ are Fourier modes defined with respect to the ``rod'' scalar field coordinates $\phi^i$.

The relative strength of these quantum fluctuations when compared to the background expectation value $V(\phi^J)$ depends on the specific functional form of the mean field, even within the approximations made. In a regime in which only a single spin $j=j_0$ contributes, one could simplify the previous expression to
\ben
\langle \delta \hat{V}(\phi^0,K^i)\delta \hat{V}(\phi^0,K'^i)\rangle = (2\pi)^3 \delta(K^i+K'^i)\,V_{j}\,V(\phi^0)
\label{vcorr}
\een
so that there is no longer an explicit dependence on the mean field. In this simple case, we see explicitly that the amplitude of these fluctuations scales as $\sqrt{V(\phi^0)}$, as one would expect in general for extensive quantities such as the volume.

\subsection{Towards a Cosmological GFT Condensate Scenario: Gauge-Invariant Perturbations}
\label{gaugeinvperp}
\begin{figure}[htp]
\includegraphics[scale=0.85]{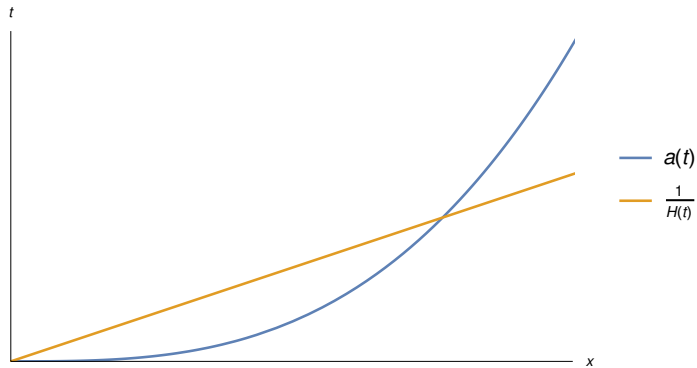}
\caption{Scale factor $a(t)$ vs. evolution of the Hubble radius in the classical cosmology without a bounce.}
\label{fig2}
\end{figure}

We now consider a cosmological scenario in which the expanding universe of classical cosmology is preceded by a GFT condensate phase of quantum gravity origin, which provides both an initial state for the background evolution and for perturbations on top of it. Recall that in section \ref{sec2} the need for a bounce and previous collapsing phase, for a universe dominated a massless scalar field, was to provide a physical mechanism for the generation of cosmological perturbations; the perturbations arose as vacuum fluctuations inside the Hubble radius, then left the Hubble radius in the contracting phase in order to re-enter after the bounce. A scenario only based on a massless scalar cosmology in classical general relativity is not sufficient (Fig.~\ref{fig2}): in such a scenario, for every mode there is only the possibility to enter, never to leave the Hubble radius. Perturbations seen today must have originated outside the Hubble radius, where there is no causal mechanism to generate them.

In the GFT condensate scenario we envisage, the solution to this problem is to postulate a prior quantum gravity phase in which volume perturbations exist as quantum fluctuations, with two-point function given by (\ref{volumeperps}). That is, perturbations are generated with constant amplitude on all scales as given by the Fourier modes $K^i$. The notion of a classical Hubble radius is not well defined in this deep quantum-gravity phase since any notion of geometry is subject to large quantum fluctuations, and large deviations from the notion of causality as given by classical general relativity are, at least a priori, acceptable. The point of this section is to extract the implications of (\ref{volumeperps}) for the initial state of perturbations in the expanding phase.

Consider the notion of local volume perturbation in conventional cosmology. If we parametrise scalar perturbations in the metric by a (Bardeen) variable $\Phi$ in longitudinal gauge,
\ben
{\rm d}s^2 = -a(\eta)^2(1+2\Phi(\eta,x)){\rm d}\eta^2 + a(\eta)^2\,(1-2\Phi(\eta,x))h_{ij}(x){\rm d}x^i{\rm d}x^j\,,
\een
a local volume perturbation would be given by
\ben
\delta V(\eta,x)=-3a^3(\eta)\Phi(\eta,x)\,.
\een
Its correlation function would then be
\ben
\langle \delta V(\eta,k)\delta V(\eta,k') \rangle = 9 a^6(\eta)\langle \Phi(\eta,k)\Phi(\eta,k') \rangle = \frac{144\pi^5}{k^3}\,a^6(\eta)\,\Delta^2_\Phi(\eta,k)\,\delta(k+k')
\een
with $\Delta^2_\Phi$ the (dimensionless) power spectrum for the metric perturbation $\Phi$, again evaluated in longitudinal gauge. In general, the definition of a volume perturbation is gauge-dependent and its relation to the curvature perturbation variable
\ben
\zeta = \Phi + \frac{\delta\rho}{{\rm d\rho}/{\rm d}\eta}\frac{1}{a}\frac{{\rm d}a}{{\rm d}\eta}
\label{curvpert}
\een
is then also gauge-dependent. It is the latter that we are really interested in. Hence, as already discussed in \cite{GFCperp}, we need to go beyond volume perturbations and also consider fluctuations in the matter density in order to connect to the relevant power spectrum in $\zeta$.

The calculations presented for volume perturbations in GFT condensates can be straightforwardly generalised to any (extensive) observable of the form
\ben
\delta \hat{\mathcal{O}}(\phi^J) = \hat{\mathcal{O}}(\phi^J) - \langle\hat{\mathcal{O}}(\phi^J)\rangle\,,\quad \hat{\mathcal{O}}(\phi^J)\equiv \sum_{\{j_I\}} \left[\hat\varphi^\dagger_{\{j_I\}}(\phi^J)\mathcal{O}_{\{j_I\}} \hat\varphi_{\{j_I\}}(\phi^J)+\left(\mathcal{O}_{\{j_I\}}\hat\varphi_{\{j_I\}}(\phi^J)\right)^\dagger \hat\varphi_{\{j_I\}}(\phi^J)\right]
\een
where we have added a Hermitian conjugate to ensure that $\hat{\mathcal{O}}(\phi^J)$ is Hermitian. Focusing specifically on the matter density, we first define a kinetic energy density in each of the four $\phi^I$ by
\ben
\rho^I(\phi^J)=\half\left(\frac{\langle\hat\pi_\phi^I(\phi^J)\rangle}{\langle\hat{V}(\phi^J)\rangle}\right)^2
\label{kineticenergy}
\een
where $\hat\pi_\phi^I(\phi^J)$, the canonical momentum of the scalar field $\phi^I$, is the observable obtained by setting $\mathcal{O}_{\{j_I\}}\equiv-\frac{\im}{2}\frac{\partial}{\partial\phi^I}$ \cite{GFTfriedmann}. In (\ref{kineticenergy}), we have replaced classical expressions for the kinetic energy density by expectation values in GFT. Perturbations in the total energy density $\rho(\phi^J)$ are then given by
\ben
\delta\rho(\phi^J)=\sum_I \frac{\pi_\phi^I(\phi^J)\delta\pi_\phi^I(\phi^J)}{V^2(\phi^J)}-2\rho(\phi^J)\frac{\delta V(\phi^J)}{V(\phi^J)}\,.
\label{energydensity}
\een
We neglect gradient energy since the dynamics of free scalar fields is completely dominated by time derivatives, i.e., their kinetic energy, near a classical singularity; this is essentially the statement of the BKL conjecture in this context (see, e.g., \cite{BLKbook}). Then notice that for a homogeneous, isotropic mean field of the form $\sigma_{\{j_I\}}(\phi^J)=\sigma_j(\phi^0)$ we have
\ben
\langle \hat\pi_\phi^i(\phi^J)\rangle = 0\,,\quad i=1,2,3
\label{pivanish}
\een
and thus only the clock field $\phi^0$ contributes non-negligibly to (\ref{energydensity}). This implies we only need to consider correlation functions of $\delta\pi_\phi^0(\phi^J)$ and $\delta V(\phi^J)$, and using all commutation relations as defined in section \ref{volumeperp} these are easy to evaluate:
\bena
\langle \delta \hat{\pi}_\phi^0(\phi^0,\phi^i)\delta \hat{\pi}_\phi^0(\phi^0,\phi'^i)\rangle &=& \delta(\phi^i-\phi'^i)\sum_j \left|\half\sigma'_j(\phi^0)\right|^2\,,
\label{picorr}
\\\langle \delta \hat{\pi}_\phi^0(\phi^0,\phi^i)\delta \hat V(\phi^0,\phi'^i)\rangle &=& \frac{\im}{2}\delta(\phi^i-\phi'^i)\sum_j V_j\bar\sigma'_j(\phi^0)\sigma_j(\phi^0)\,.
\label{mixcorr}
\eena
The property of volume perturbations that the right-hand side is independent of the $\phi^i$ apart from an overall delta function carries through to these quantities, by virtue of the assumed homogeneity of the mean field. It is then clear that the effective curvature perturbation variable $\zeta$ resulting from the initial fluctuations in the GFT condensate satisfies
\ben
\langle\zeta(k)\zeta(k')\rangle_{{\rm GFT}} = (2\pi)^3 \delta(k+k')\,\mathcal{A}_{{\rm GFT}}
\een
for a $k$-independent amplitude $\mathcal{A}$, i.e., the dimensionless power spectrum scales as $k^3$ and the resulting spectral index is $n_s=4$, just as for the semiclassical bounce cosmology in section \ref{sec2}. We now calculate this in detail to get some insight on the form of $\mathcal{A}$. For concreteness and simplicity, we again assume that only a single spin component $j=j_0$ contributes significantly to the mean field, in which case one can eliminate $\sigma_j$ from the expressions in favour of cosmological variables.

From the definition (\ref{curvpert}) of the curvature perturbation variable and with $\Phi\rightarrow -\frac{\delta V}{3V}$, we have
\ben
\langle \zeta(k)\zeta(k')\rangle = \frac{1}{9V^2}\langle \delta V(k)\delta V(k')\rangle -\frac{2}{3V}\frac{a'}{\rho'\,a}\langle \delta V(k)\delta \rho(k')\rangle+\left(\frac{a'}{\rho'\,a}\right)^2\langle \delta \rho(k)\delta \rho(k')\rangle
\een
Notice that the combination $\frac{a'}{\rho'}\equiv \frac{{\rm d}a/{\rm d}\eta}{{\rm d}\rho/{\rm d}\eta}$ is reparametrisation invariant with respect to a change of time variable from $\eta$ to, e.g., our scalar field time variable $\phi^0$, so it can be matched with the corresponding expression in GFT. For $\delta\rho$ we can now use (\ref{energydensity}), taking into account that only a single $\pi_\phi\equiv\pi_\phi^0$, the momentum conjugate to the clock scalar $\phi^0$, is non-vanishing. We also replace the scale factor $a$ by $a\propto V^{1/3}$. After some simple algebra, we obtain
\ben
\langle \zeta(k)\zeta(k')\rangle = \frac{1}{9\pi_\phi^2}\langle \delta \pi_\phi(k)\delta \pi_\phi(k')\rangle
\label{zetacorr}
\een
where we now write $\pi_\phi$ for $\pi_\phi^0$ due to (\ref{pivanish}). We see that only perturbations in the scalar field momentum $\pi_\phi$ are relevant for the spectrum of the perturbation variable $\zeta$. This is true in classical cosmology as well as in GFT condensate cosmology.\footnote{The careful reader will have noticed an important technical point. In replacing the metric perturbation variable $\Phi$ directly by a volume perturbation, we assume that in the general form of a metric containing scalar perturbations, \[{\rm d}s^2 = -a(\eta)^2(1+2\Phi(\eta,x)){\rm d}\eta^2 + 2a(t)\,\partial_i B(\eta,x)\,{\rm d}x^i{\rm d}\eta+ a(t)^2\,(1-2\Psi(\eta,x)+2\Delta E(\eta,x))h_{ij}(x){\rm d}x^i{\rm d}x^j\,,\]we can choose a gauge in which $\Delta E=0$. ($\Phi=\Psi$ follows from the absence of anisotropic stress.) In the setting of GFT condensate cosmology, one always works in a harmonic gauge \cite{harmonic}, and longitudinal gauge with $\Delta E=0$ everywhere is not a possible choice. What {\em is} however possible within harmonic gauge is to set $\Delta E=0$ on an initial hypersurface on which perturbations are generated \cite{Battarra:2014tga}, which is sufficient for our purposes.} We can now use the correlation function (\ref{picorr}), in the simplest case where only a single spin contributes, to compute an explicit expression for the amplitude $\mathcal{A}$. Notice that, as we have already emphasised, the right-hand side of (\ref{zetacorr}) results in a spectral index $n_s=4$ since all two-point functions are independent of the $\phi^i$.

In order to relate (\ref{picorr}) to cosmological observables, we observe that for a single-spin condensate (i.e., a mean field whose only non-vanishing contribution is for $j=j_0$),
\ben
\left|\half\sigma'(\phi^0)\right|^2=\frac{1}{16}\frac{V'(\phi^0)^2}{V_{j_0}V(\phi^0)} + \frac{1}{4}\pi_\phi^2(\phi^0) \frac{V_{j_0}}{V(\phi^0)}
\een
with $\pi_\phi(\phi^0)\equiv -\frac{\im}{2}(\bar\sigma(\phi^0)\sigma'(\phi^0)-\bar\sigma'(\phi^0)\sigma(\phi^0))$. Putting it all together, we then find
\ben
\mathcal{A}_{{\rm GFT}}(\phi^0_*) = \frac{V_{j_0}}{36V}+\frac{(V')^2}{144 V_{j_0}\pi_\phi^2V}\,.
\label{amplitude}
\een
Here $V\equiv V(\phi^0_*)$ corresponds to the value of the local volume element (corresponding to the scale factor cubed) evaluated at a scalar field clock time $\phi^0=\phi^0_*$ where the spectrum is generated. The dimensionless power spectrum for $\zeta$ is given by
\ben
\Delta_\zeta^2(k)=\frac{k^3}{2\pi^2}\mathcal{A}_{{\rm GFT}}(\phi^0_*,k)\,.
\een
In general, one should expect the expression (\ref{amplitude}) for the amplitude to be dominated by the second term:  the first term is of order $m_{{\rm Pl}}^{-3}$ ($V_{j_0}$ is the elementary volume associated to each ``building block'' of the GFT condensate), but the second one is of order $m_{{\rm Pl}}/M^4$ where $M$ is the mass scale set by the energy density in the scalar field (see~(\ref{energydens})), and we assume $m_{{\rm Pl}}> M$. We then obtain the leading order result (using, from (\ref{classvol}), $V'\approx\sqrt{12\pi G}V$)
\ben
\Delta_\zeta^2(k)\sim\frac{ m_{{\rm Pl}} V}{24\pi j_0^{3/2} M^4}k^3\,.
\een
As we have stressed before, this result is consistent with the calculation in semiclassical cosmology as far as the spectral index $n_s=4$ is concerned. It is remarkable that a pure quantum gravity calculation, assuming that the origin of cosmological perturbations is in the quantum fluctuations of a quantum gravity condensate, yields a spectrum consistent with the expectations from semiclassical physics. This agreement is the main result of this paper.

The point of presenting the details of the GFT calculation was to also obtain insights about the form of the amplitude, which is different from that in the semiclassical theory. Writing the resulting amplitude in a similar form as we did in (\ref{classiAmpl}) for the semiclassical case, we have
\ben
\Delta_\zeta^2(k)\sim 0.01 \left(\frac{ m_{{\rm Pl}}^2 V}{j_0^{3/2} M^2}\right)\frac{k^3}{M^2 m_{{\rm Pl}}}
\label{GFTampl}
\een
where the terms in brackets give an additional factor contributing to the amplitude, replacing the expression $\log\left(\left(\frac{2\pi G}{3}\right)^{3/4} 32\sqrt{2\rho_c}\,Mk^{-3/2}\right)^2$ we found in the semiclassical calculation. With $m_{{\rm Pl}}> M$ and $j_0$ of order one in most models, one might expect that
\ben
\frac{ m_{{\rm Pl}}^2 V}{j_0^{3/2} M^2}\gg 1\,;
\een
however, $V$ is a local volume element, related to the total volume of the condensate by
\ben
V_{{\rm tot}}(\phi^0)=\int {\rm d}^3\phi\;V(\phi^0,\phi^i)\,.
\een
For a homogeneous function $V(\phi^0,\phi^i)\equiv V(\phi^0)$, which is the case we have been discussing, $V$ is the ratio of the physical condensate volume and the coordinate volume given the allowed values of $\phi^i$. So far we did not worry about the regularisation of the total volume, which is infinite for a non-compact homogeneous universe and must be regularised. In standard cosmology, while the source of some technical complications, this point is of relatively minor importance as one can work in a finite box and at the end take the limit in which finite-size effects vanish. In GFT condensates, however, this point is more important, since the particle number in the condensate cannot become too large before interactions dominate \cite{GFTfriedmann}. Thus, the total volume $V_{{\rm tot}}$ is bounded from above. The limit of large coordinate volume would correspond to simultaneously taking $V(\phi^0)\rightarrow 0$. We must hence work in a finite box for the allowed coordinate values $\phi^i$ and with finite, small $V$.

The difference between the semiclassical prediction for the amplitude (\ref{classiAmpl}) and the GFT result (\ref{GFTampl}) is then a combination of $m_{{\rm Pl}}$ and effectively three free parameters: a mass scale $M<m_{{\rm Pl}}$ associated to the matter energy density; a spin $j_0$ (a number of order one) corresponding to the (in this simple case single) volume eigenstate occupied by quanta in the GFT condensate; and a volume factor $V\ll 1$ relating the coordinate volume to the physical volume in the condensate (bounded from above since we assumed that the free GFT dynamics provide a good approximation). Out of these, the last two can be derived or at least constrained from a more specific choice of GFT dynamics, which will set a precise upper bound on the total condensate volume, and single out $j_0$ as the spin for which a ratio of couplings takes its maximum (see the discussion before (\ref{volumeeq})). $M$ will presumably remain as a free parameter, like in the semiclassical theory. From a phenomenological perspective, then, this model appears to be as predictive as the semiclassical cosmology of a bouncing universe filled with a massless scalar field.\footnote{The critical density $\rho_c$ in LQC, which appears in the logarithmic contribution to the amplitude, has a similar status as $j_0$ for GFT condensates: $\rho_c$ depends on the details of the assumed loop quantum gravity state \cite{BenAchour:2016ajk}.} From a fundamental viewpoint, however, the results are promising for the status of GFT condensates as a quantum gravity origin for the emergence of a semiclassical universe: not only the homogeneous background dynamics, but also the physics of perturbations give results compatible with semiclassical theory. It is far from obvious that a background-independent approach to quantum gravity in which space and time emerge from ``pre-geometric'' discrete fundamental degrees of freedom should achieve this.

\section{Discussion}
\label{sec4}

Showing the emergence of a realistic universe from a theory of quantum gravity is an outstanding challenge for fundamental physics, addressing which will be crucial in order to complete our understanding of the origins of space and time. In this article, we showed some progress towards addressing it in the group field theory approach to quantum gravity: we showed how a condensate of discrete ``building blocks'' of geometry, a candidate for a semiclassical macroscopic geometry, is capable of producing not only the correct dynamics of an exactly homogeneous and isotropic background spacetime, but also of generating a power spectrum of primordial scalar perturbations that is consistent with expectations coming from the semiclassical approach of standard cosmology in terms of quantum fields on a curved spacetime background. For technical reasons, the matter content was restricted to free, massless scalar fields and the background spacetime was spatially flat. This is because this is the only model that has so far been constructed within GFT \cite{GFTfriedmann}. While not viable as a model for our own Universe as far as the matter content is concerned, this is also a toy model on which much work in quantum cosmology, from loop quantum cosmology \cite{LQC} to a variety of other approaches \cite{BlythIsham,Hoehn}, has been based, and thus the results obtained here should facilitate a more detailed comparison with all these approaches regarding methods, technical questions, and physical predictions.

In the GFT condensate cosmology scenario that this article is beginning to develop, the origin of cosmological perturbations is neither in an expanding inflationary phase nor in a collapsing phase, as in usual bounce models \cite{bounceReview}; it is in the quantum fluctuations of the GFT condensate making up space, time and matter itself at the fundamental level. From the perspective of low-energy, semiclassical cosmology, this mechanism appears to violate causality as it seems to require generation of perturbation modes outside of the Hubble radius. However, from the GFT (or generally, quantum gravitational) perspective this is not the right way of looking at things: here, space and time are themselves made up of quantum gravitational degrees of freedom, with necessary fluctuations in the effective spacetime geometry due to the uncertainty principle. {\em Any} question involving spacetime geometry must then come with an intrinsic uncertainty, and this pertains to questions such as whether physical correlations respect causality, since the notion of causality is itself only defined once a metric is known. Such conundrums are of course known to practitioners of quantum gravity, and have been discussed since the early days of the subject; the new input they might lend to fundamental cosmology is in a revision of our usual concepts of causality applied to the very early universe in the Planckian regime. In spirit at least, our work has some relation to the proposal of \cite{joemijoe} for generating a scale-invariant spectrum of cosmological perturbations in a non-geometric ``high temperature phase'', where the concept of causality is entirely undefined (and instead only the holographic principle is used).

Of course, in our model the power spectrum does not come out to be (nearly) scale-invariant as required by what we see in the sky, but instead gives a spectral index $n_s=4$, as one would expect for a universe filled with a free scalar field (with matter equation of state $P=\rho$) by simply considering the order of the Hankel functions solving the Mukhanov-Sasaki equation (see, e.g., \cite{brandenfinelli,wands}). A major challenge for GFT condensate cosmology would be the incorporation of more complicated matter dynamics that may reproduce an approximately scale-invariant, and hence observationally viable, power spectrum. The obvious candidate for this is the conventional matter bounce scenario \cite{brandenfinelli}, i.e., a universe dominated by dust, as discussed within LQC in \cite{edmatterbounce}. Since the effective dynamics for GFT condensates is derived from a fundamental GFT action for quantum gravity coupled to suitable matter, this means that the GFT formalism itself must be extended from describing free, massless scalars, as in \cite{lioritizhang}, to at least self-interacting scalar fields with nontrivial potential terms. The details of the resulting cosmological power spectrum would also be affected by generalising the calculation we presented, e.g., by including the effect of GFT condensate interactions or relaxing the assumption that the mean field configuration is exactly homogeneous and isotropic.

It is interesting that, in contrast to the usual formalism of quantum field theory on curved spacetime which requires a choice of vacuum, here only the general statement that the condensate mean field should be initially homogeneous and isotropic was needed in order to obtain the general form of the power spectrum (in particular, its spectral index). In this sense, the formalism of ``cosmology as quantum gravity hydrodynamics'' \cite{QGHydro} seems to itself provide enough constraints to deduce general features of the spectrum of perturbations, as already observed in \cite{GFCperp}. The underlying explanation of this initially puzzling statement is that the dynamics of spacetime and matter are more intricately linked in a full quantum gravity setting, where there is only a single quantum state for both. More precisely, demanding homogeneity for the background spacetime, by fixing the general form of the condensate mean field, then also fixed all higher order $n$-point functions including the ones responsible for perturbations. The latter statement is a general property (in some sense the defining property) of the mean-field approximation we are working in; hence a departure from this strict connection between background and perturbations could also be implied by a breakdown of the mean-field approximation itself. Understanding the range of validity of the mean-field approximation within GFT condensates is one of the main open questions in this research programme; however, this is a question within a specific quantum field theory (namely GFT), disconnected from any general or conceptual discussion of initial conditions in cosmology. This article outlines a direct avenue in which such theoretical questions within quantum gravity can have a direct link to cosmological phenomenology. Overall, while still far from providing a satisfactory description of the observed cosmology, the fact that in GFT condensate cosmology we are now able to show such connections in explicit calculations is a major step forward towards connecting background-independent quantum gravity to reality.

{\em Acknowledgements.} --- I would like to thank Daniele Oriti, Axel Polaczek and Edward Wilson-Ewing for helpful comments on earlier versions of the manuscript, and Edward Wilson-Ewing for comments related to \cite{GFCperp} that partially motivated this analysis. The research leading to these results was funded by the Royal Society under a Royal Society University Research Fellowship (UF160622) and a Research Grant for Research Fellows (RGF\textbackslash R1\textbackslash 180030).


\begin{thebibliography}{999}
\bibitem{Baumann:2009ds} D.~Baumann, Inflation. In \doin{10.1142/9789814327183_0010}{{\em Physics of the Large and Small, TASI 2009, Proceedings of the Theoretical}} \doin{10.1142/9789814327183_0010}{{\em Advanced Study Institute in Elementary Particle Physics}}, eds. C.~Csaki, S.~Dodelson (World Scientific, 2011), pp. 523-686, \arX{0907.5424}.
\bibitem{bounceReview}  R.~Brandenberger and P.~Peter, Bouncing Cosmologies: Progress and Problems. \doin{10.1007/s10701-016-0057-0}{{\em Found.\ Phys.} {\bf 47}} \doin{10.1007/s10701-016-0057-0}{(2017), 797-850}, \arX{1603.05834}.
\bibitem{TransPl} J.~Martin and R.~H.~Brandenberger, Trans-Planckian problem of inflationary cosmology. \doin{10.1103/PhysRevD.63.123501}{{\em Phys.\ Rev.\ D}} \doin{10.1103/PhysRevD.63.123501}{{\bf 63} (2001), 123501}, \oarX{hep-th/0005209}.
\bibitem{noboundary}  J.~B.~Hartle and S.~W.~Hawking, Wave function of the Universe. \doin{10.1103/PhysRevD.28.2960}{{\em Phys.\ Rev.\ D} {\bf 28} (1983), 2960-2975}.
\bibitem{controversy} J.~Feldbrugge, J.~L.~Lehners, and N.~Turok, No Smooth Beginning for Spacetime. \doin{10.1103/PhysRevLett.119.171301}{{\em Phys.\ Rev.\ Lett.}} \doin{10.1103/PhysRevLett.119.171301}{{\bf 119} (2017),  171301}, \arX{1705.00192};
\\  J.~Diaz Dorronsoro, J.~J.~Halliwell, J.~B.~Hartle, T.~Hertog, O.~Janssen, and Y.~Vreys, Damped Perturbations in the No-Boundary State. \doin{10.1103/PhysRevLett.121.081302}{{\em Phys.\ Rev.\ Lett.}  {\bf 121} (2018),  081302}, \arX{1804.01102};
\\  J.~Feldbrugge, J.~L.~Lehners, and N.~Turok, Inconsistencies of the New No-Boundary Proposal. \doin{10.3390/universe4100100}{{\em Universe} {\bf 4} (2018),  100}, \arX{1805.01609}.
\bibitem{brandenfinelli}   F.~Finelli and R.~Brandenberger, Generation of a scale-invariant spectrum of adiabatic fluctuations in cosmological models with a contracting phase. \doin{10.1103/PhysRevD.65.103522}{{\em Phys.\ Rev.\ D} {\bf 65} (2002), 103522}, \oarX{hep-th/0112249}.
\bibitem{qprop}  S.~Gielen and N.~Turok, Quantum propagation across cosmological singularities. \doin{10.1103/PhysRevD.95.103510}{{\em Phys.\ Rev.\ D} {\bf 95}} \doin{10.1103/PhysRevD.95.103510}{(2017),  103510}, \arX{1612.02792}.
\bibitem{philo} D.~Oriti, Disappearance and emergence of space and time in quantum gravity. \doin{10.1016/j.shpsb.2013.10.006}{{\em Stud.\ Hist.\ Phil.}} \doin{10.1016/j.shpsb.2013.10.006}{{\em Sci.\ B} {\bf 46} (2014), 186-199}, \arX{1302.2849}.
\bibitem{GFTreviews}    D.~Oriti, The microscopic dynamics of quantum space as a group field theory. In {\em Foundations of Space and Time: Reflections on Quantum Gravity}, eds. G.~Ellis, J.~Murugan, A.~Weltman (Cambridge University Press, 2012), pp. 257-320, \arX{1110.5606};
\\A. Baratin and D. Oriti, Ten questions on Group Field Theory (and their tentative answers). \doin{10.1088/1742-6596/360/1/012002}{{\em J.\ Phys.\ Conf.\ Ser.}  {\bf 360} (2012), 012002}, \arX{1112.3270};
\\T. Krajewski, Group Field Theories. \doin{10.22323/1.140.0005}{{\em PoS QGQGS} {\bf 2011}, 005}, \arX{1210.6257}.
\bibitem{matrixtensormodels} P.~Di Francesco, P.~H.~Ginsparg, and J.~Zinn-Justin, 2D gravity and random matrices. \doin{10.1016/0370-1573(94)00084-G}{{\em Phys.\ Rept.}} \doin{10.1016/0370-1573(94)00084-G}{{\bf 254} (1995) 1-133}, \oarX{hep-th/9306153};
\\R.~Gurau and J.~P.~Ryan, Colored Tensor Models - a Review. \doin{10.3842/SIGMA.2012.020}{{\em SIGMA} {\bf 8} (2012), 020}, \arX{1109.4812};
\\R.~Gurau, {\em Random Tensors} (Oxford University Press, 2017).
\bibitem{LQG} A.~Ashtekar and J.~Pullin (eds.), {\em Loop Quantum Gravity: The First 30 Years (100 Years of General Relativity: Volume 4)} (World Scientific, Singapore, 2017).
\bibitem{GFT2ndq}  D.~Oriti, Group field theory as the second quantization of loop quantum gravity. \doin{10.1088/0264-9381/33/8/085005}{{\em Class.\ Quant.\ Grav.}} \doin{10.1088/0264-9381/33/8/085005}{{\bf 33} (2016),  085005},  \arX{1310.7786}.
\bibitem{GFTcosmo}  S.~Gielen, D.~Oriti, and L.~Sindoni, Cosmology from Group Field Theory Formalism for Quantum Gravity, \doin{10.1103/PhysRevLett.111.031301}{{\em Phys.\ Rev.\ Lett.}  {\bf 111} (2013),  031301}, \arX{1303.3576};
\\S.~Gielen, D.~Oriti, and L.~Sindoni, Homogeneous cosmologies as group field theory condensates. \doin{10.1007/JHEP06(2014)013}{{\em JHEP} {\bf 1406} (2014), 013}, \arX{1311.1238}.
\bibitem{GFTfriedmann}   D.~Oriti, L.~Sindoni, and E.~Wilson-Ewing, Emergent Friedmann dynamics with a quantum bounce from quantum gravity condensates. \doin{10.1088/0264-9381/33/22/224001}{{\em Class.\ Quant.\ Grav.}  {\bf 33} (2016),  224001}, \arX{1602.05881};
\\ D.~Oriti, L.~Sindoni, and E.~Wilson-Ewing, Bouncing cosmologies from quantum gravity condensates. \doin{10.1088/1361-6382/aa549a}{{\em Class.\ Quant.\ Grav.} {\bf 34} (2017),  04LT01}, \arX{1602.08271}.\bibitem{BlythIsham}  W.~F.~Blyth and C.~J.~Isham, Quantization of a Friedmann universe filled with a scalar field. \doin{doi:10.1103/PhysRevD.11.768}{{\em Phys.\ Rev.\ D} {\bf 11} (1975), 768-778}.
\bibitem{LQC}  A.~Ashtekar and P.~Singh, Loop quantum cosmology: a status report. \doin{10.1088/0264-9381/28/21/213001}{{\em Class.\ Quant.\ Grav.}  {\bf 28} (2011),} \doin{10.1088/0264-9381/28/21/213001}{213001}, \arX{1108.0893};
\\K.~Banerjee, G.~Calcagni, and M.~Martin-Benito, Introduction to Loop Quantum Cosmology. \doin{10.3842/SIGMA.2012.016}{{\em SIGMA} {\bf 8}} \doin{10.3842/SIGMA.2012.016}{(2012), 016}, \arX{1109.6801};
\\M.~Bojowald, Loop quantum cosmology. {\em Living Rev.\ Rel.}  {\bf 11} (2008), 4.
\bibitem{GFCreview} S.~Gielen and L.~Sindoni, Quantum Cosmology from Group Field Theory Condensates: a Review. \doin{10.3842/SIGMA.2016.082}{{\em SIGMA} {\bf 12} (2016), 082}, \arX{1602.08104};
\\D.~Oriti, The universe as a quantum gravity condensate. \doin{10.1016/j.crhy.2017.02.003}{{\em Comptes Rendus Physique} {\bf 18} (2017), 235}, \arX{1612.09521}.
\bibitem{improdyn} A.~Ashtekar, T.~Pawlowski, and P.~Singh, Quantum nature of the big bang: Improved dynamics. \doin{10.1103/PhysRevD.74.084003}{{\em Phys.\ Rev.\ D} {\bf 74} (2006), 084003}, \oarX{gr-qc/0607039}.
\bibitem{LQGLQC}  E.~Alesci and F.~Cianfrani, Loop quantum cosmology from quantum reduced loop gravity. \doin{10.1209/0295-5075/111/40002}{{\em EPL}} \doin{10.1209/0295-5075/111/40002}{{\bf 111} (2015),  40002}, \arX{1410.4788};
\\ E.~Alesci and F.~Cianfrani, Quantum reduced loop gravity and the foundation of loop quantum cosmology. \doin{10.1142/S0218271816420050}{{\em Int.\ J.\ Mod.\ Phys.\ D} {\bf 25} (2016),  1642005}, \arX{1602.05475};
\\ N.~Bodendorfer, An embedding of loop quantum cosmology in $(b,v)$ variables into a full theory context. \doin{10.1088/0264-9381/33/12/125014}{{\em Class.\ Quant.\ Grav.}  {\bf 33} (2016),  125014}, \arX{1512.00713}.
\bibitem{GFCperp} S.~Gielen and D.~Oriti, Cosmological perturbations from full quantum gravity. \doin{10.1103/PhysRevD.98.106019}{{\em Phys.\ Rev.\ D} {\bf 98}} \doin{10.1103/PhysRevD.98.106019}{(2018), 106019}, \arX{1709.01095}.
\bibitem{harmonic}   S.~Gielen, Group field theory and its cosmology in a matter reference frame. \doin{10.3390/universe4100103}{{\em Universe} {\bf 4} (2018),  103}, \arX{1808.10469}.
\bibitem{matterbounce}   R.~H.~Brandenberger, Alternatives to the inflationary paradigm of structure formation. \doin{10.1142/S2010194511000109}{{\em Int.\ J.\ Mod.}} \doin{10.1142/S2010194511000109}{{\em Phys.\ Conf.\ Ser.}  {\bf 01} (2011), 67-79}, \arX{0902.4731}.
\bibitem{edmatterbounce}  E.~Wilson-Ewing, The Matter Bounce Scenario in Loop Quantum Cosmology. \doin{10.1088/1475-7516/2013/03/026}{{\em JCAP} {\bf 1303} (2013), 026}, \arX{1211.6269}.
\bibitem{stringGas}   R.~H.~Brandenberger, String gas cosmology after Planck, \doin{doi:10.1088/0264-9381/32/23/234002}{{\em Class.\ Quant.\ Grav.}  {\bf 32} (2015),  234002}, \arX{1505.02381}.
\bibitem{taveras}   V.~Taveras, Corrections to the Friedmann equations from loop quantum gravity for a universe with a free scalar field. \doin{10.1103/PhysRevD.78.064072}{{\em Phys.\ Rev.\ D} {\bf 78} (2008), 064072}, \arX{0807.3325}.
\bibitem{BrandenbergerIntro} R.~H.~Brandenberger, Beyond Standard Inflationary Cosmology. To appear in {\em Beyond Spacetime: The Foundations of Quantum Gravity}, eds. N.~Huggett, K.~Matsubara, and C.~Wuethrich (Cambridge University Press, 2018), \arX{1809.04926}.
\bibitem{sasakimukhanov}   M.~Sasaki, Large Scale Quantum Fluctuations in the Inflationary Universe. \doin{10.1143/PTP.76.1036}{{\em Prog.\ Theor.\ Phys.}} \doin{10.1143/PTP.76.1036}{{\bf 76} (1986) 1036-1046};
\\V.~F.~Mukhanov, Quantum Theory of Gauge Invariant Cosmological Perturbations. {\em Sov.\ Phys.\ JETP} {\bf 67} (1988) 1297-1302 [{\em Zh.\ Eksp.\ Teor.\ Fiz.}  {\bf 94N7} (1988) 1-11].
\bibitem{LQCpert} T.~Cailleteau, J.~Mielczarek, A.~Barrau, and J.~Grain, Anomaly-free scalar perturbations with holonomy corrections in loop quantum cosmology. \doin{10.1088/0264-9381/29/9/095010}{{\em Class.\ Quant.\ Grav.}  {\bf 29} (2012), 095010}, \arX{1111.3535};
\\  T.~Cailleteau and A.~Barrau, Gauge invariance in loop quantum cosmology: Hamilton-Jacobi and Mukhanov-Sasaki equations for scalar perturbations. \doin{10.1103/PhysRevD.85.123534}{{\em Phys.\ Rev.\ D} {\bf 85} (2012), 123534}, \arX{1111.7192}.
\bibitem{BCGFT}  R.~De Pietri, L.~Freidel, K.~Krasnov, and C.~Rovelli, Barrett-Crane model from a Boulatov-Ooguri field theory over a homogeneous space. \doin{10.1016/S0550-3213(00)00005-5}{{\em Nucl.\ Phys.\ B} {\bf 574} (2000) 785-806}, \oarX{hep-th/9907154}.
\bibitem{barrettcrane}  J.~W.~Barrett and L.~Crane, Relativistic spin networks and quantum gravity. \doin{10.1063/1.532254}{{\em J.\ Math.\ Phys.}} \doin{10.1063/1.532254}{{\bf 39} (1998) 3296-3302}, \oarX{gr-qc/9709028};
\\J.~W.~Barrett and L.~Crane, A Lorentzian signature model for quantum general relativity. \doin{10.1088/0264-9381/17/16/302}{{\em Class.}} \doin{10.1088/0264-9381/17/16/302}{{\em Quant.\ Grav.}  {\bf 17} (2000) 3101-3118}, \oarX{gr-qc/9904025}.
\bibitem{rovellireis}   M.~P.~Reisenberger and C.~Rovelli, Spacetime as a Feynman diagram: the connection formulation. \doin{10.1088/0264-9381/18/1/308}{{\em Class.\ Quant.\ Grav.}  {\bf 18} (2001) 121-140}, \oarX{gr-qc/0002095}.
\bibitem{GFTrenorm}
J. Ben Geloun and V. Bonzom, Radiative Corrections in the Boulatov-Ooguri Tensor Model: The 2-Point Function. \doin{10.1007/s10773-011-0782-2}{{\em Int.\ J.\ Theor.\ Phys.}  {\bf 50} (2011) 2819-2841}, \arX{1101.4294};
\\S. Carrozza, D. Oriti, and V. Rivasseau, Renormalization of a $SU(2)$ Tensorial Group Field Theory in Three Dimensions. \doin{10.1007/s00220-014-1928-x}{{\em Commun.\ Math.\ Phys.}  {\bf 330} (2014) 581-637}, \arX{1303.6772}.
\bibitem{AshtekarBarbero}   A.~Ashtekar, New Variables for Classical and Quantum Gravity. \doin{10.1103/PhysRevLett.57.2244}{{\em Phys.\ Rev.\ Lett.}  {\bf 57} (1986) 2244-2247};
\\  A.~Ashtekar, New Hamiltonian formulation of general relativity. \doin{10.1103/PhysRevD.36.1587}{{\em Phys.\ Rev.\ D} {\bf 36} (1987) 1587-1602};
\\  J.~F.~Barbero G., Real Ashtekar variables for Lorentzian signature space-times. \doin{10.1103/PhysRevD.51.5507}{{\em Phys.\ Rev.\ D} {\bf 51}} \doin{10.1103/PhysRevD.51.5507}{{\bf 51} (1995) 5507-5510}, \oarX{gr-qc/9410014}.
\bibitem{lioritizhang} Y.~Li, D.~Oriti, and M.~Zhang, Group field theory for quantum gravity minimally coupled to a scalar field. \doin{10.1088/1361-6382/aa85d2}{{\em Class.\ Quant.\ Grav.} {\bf 34} (2017),  195001}, \arX{1701.08719}.
\bibitem{KCLgroup}
M.~de Cesare, A.G.A.~Pithis, and M.~Sakellariadou, Cosmological implications of interacting group field theory models: Cyclic Universe and accelerated expansion. \doin{10.1103/PhysRevD.94.064051}{{\em Phys.\ Rev.\ D} {\bf 94} (2016), {\em 6},  064051}, \arX{1606.00352}.
\bibitem{anisopaper}
M.~de Cesare, D.~Oriti, A.G.A.~Pithis, and M.~Sakellariadou, Dynamics of anisotropies close to a cosmological bounce in quantum gravity. \doin{10.1088/1361-6382/aa986a}{{\em Class.\ Quant.\ Grav.}~{\bf 35} (2018), {\em 1},  015014}, \arX{1709.00994}.
\bibitem{DeWitt}
B.S.~DeWitt, The Quantization of geometry. In {\em Gravitation: An introduction to current research}, ed. L.~Witten (Wiley, New York, 1962), pp. 266-381.
\bibitem{GHM}
S.B.~Giddings, D.~Marolf, and J.B.~Hartle, Observables in effective gravity. \doin{10.1103/PhysRevD.74.064018}{{\em Phys.\ Rev.\ D} {\bf 74} (2006),} \doin{10.1103/PhysRevD.74.064018}{064018}, \oarX{hep-th/0512200}.
\bibitem{sepUniv}  D.~S.~Salopek and J.~R.~Bond, Nonlinear evolution of long wavelength metric fluctuations in inflationary models, \doin{doi:10.1103/PhysRevD.42.3936}{{\em Phys.\ Rev.\ D} {\bf 42} (1990), 3936-3962};
\\D.~Wands, K.~A.~Malik, D.~H.~Lyth and A.~R.~Liddle, A New approach to the evolution of cosmological perturbations on large scales, \doin{doi:10.1103/PhysRevD.62.043527}{{\em Phys.\ Rev.\ D} {\bf 62} (2000), 043527}, \oarX{astro-ph/0003278}.
\bibitem{GFTsepUniv} F.~Gerhardt, D.~Oriti and E.~Wilson-Ewing, Separate universe framework in group field theory condensate cosmology, \doin{10.1103/PhysRevD.98.066011}{{\em Phys.\ Rev.\ D} {\bf 98} (2018),  066011}, \arX{1805.03099}.
\bibitem{bohrsommerf}  E.~Bianchi and H.~M.~Haggard, Discreteness of the Volume of Space from Bohr-Sommerfeld Quantization. \doin{10.1103/PhysRevLett.107.011301}{{\em Phys.\ Rev.\ Lett.}  {\bf 107} (2011), 011301}, \arX{1102.5439}.
\bibitem{lowj}   S.~Gielen, Emergence of a low spin phase in group field theory condensates. \doin{10.1088/0264-9381/33/22/224002}{{\em Class.\ Quant.\ Grav.}} \doin{10.1088/0264-9381/33/22/224002}{{\bf 33} (2016),  224002}, \arX{1604.06023}.
\bibitem{Toy}   E.~Adjei, S.~Gielen, and W.~Wieland, Cosmological evolution as squeezing: a toy model for group field cosmology. \doin{10.1088/1361-6382/aaba11}{{\em Class.\ Quant.\ Grav.}  {\bf 35} (2018),  105016}, \arX{1712.07266}.
\bibitem{GFThamilt}  E.~Wilson-Ewing, A relational Hamiltonian for group field theory. \arX{1810.01259}.
\bibitem{BLKbook} V.~Belinski and M.~Henneaux, {\em The Cosmological Singularity} (Cambridge University Press, 2017).
\bibitem{Battarra:2014tga} L.~Battarra, M.~Koehn, J.~L.~Lehners, and B.~A.~Ovrut, Cosmological perturbations through a non-singular ghost-condensate/Galileon bounce. \doin{10.1088/1475-7516/2014/07/007}{{\em JCAP} {\bf 1407} (2014), 007}, \arX{1404.5067}.
\bibitem{BenAchour:2016ajk}  J.~Ben Achour, S.~Brahma and M.~Geiller, New Hamiltonians for loop quantum cosmology with arbitrary spin representations. \doin{doi:10.1103/PhysRevD.95.086015}{{\em Phys.\ Rev.\ D} {\bf 95} (2017),  086015}, \arX{1612.07615}.
\bibitem{Hoehn} P.~A.~Hoehn, Switching internal times and a new perspective on the 'wave function of the universe'. \arX{1811.00611}.
\bibitem{joemijoe} J.~Magueijo, L.~Smolin, and C.~R.~Contaldi, Holography and the scale-invariance of density fluctuations. \doin{10.1088/0264-9381/24/14/009}{{\em Class.\ Quant.\ Grav.}  {\bf 24} (2007), 3691-3700}, \oarX{astro-ph/0611695}.
\bibitem{wands}  D.~Wands, Duality invariance of cosmological perturbation spectra. \doin{10.1103/PhysRevD.60.023507}{{\em Phys.\ Rev.\ D} {\bf 60} (1999), 023507}, \oarX{gr-qc/9809062}.
\bibitem{QGHydro} S.~Gielen, Identifying cosmological perturbations in group field theory condensates. \doin{10.1007/JHEP08(2015)010}{{\em JHEP} {\bf 1508}} \doin{10.1007/JHEP08(2015)010}{(2015), 010}, \arX{1505.07479}.
\end{thebibliography}
\end{document}